\documentclass[11pt]{article}

\usepackage [latin1]{inputenc}
\usepackage{ifpdf}
\usepackage{amsmath}
\usepackage{graphicx}
\usepackage{indentfirst}
\usepackage{amssymb}
\usepackage{cite}
\usepackage{color}
\usepackage{subfigure}
\usepackage[colorlinks,linkcolor=blue,citecolor=blue,urlcolor=blue,hyperindex]{hyperref}
\usepackage{indentfirst}
\usepackage{amsmath}
\usepackage{latexsym}
\usepackage{amsmath}
\usepackage{color}
\usepackage{lineno}
\usepackage{graphicx}
\usepackage{float}
\usepackage{subfigure}
\usepackage{geometry}
\usepackage{graphics}
\usepackage{subfigure}
\usepackage{graphicx}
\usepackage{multirow}
\usepackage{booktabs}
\usepackage{hyperref}
\usepackage{cite}
\usepackage{wasysym}

\definecolor{c1}{RGB}{233,202,118}
\definecolor{c2}{RGB}{220,143,1}
\definecolor{c3}{RGB}{255,107,0}
\definecolor{c4}{RGB}{255,255,255}
\definecolor{c5}{RGB}{236,213,146}
\definecolor{c6}{RGB}{227,172,60}

\begin{document}	
	\title{\textbf{Observational features of massive boson stars with thin disk accretion}}
	\date{}
	\maketitle
	\begin{center}
		\author{Guo-Ping Li}$^{a,}$$\footnote{\texttt{Corresponding author:gpliphys@yeah.net}}$,
               \author{Meng-Qi Wu}$^{a,}$$\footnote{\texttt{wumengqi202307029@163.com}}$,
        \author{Ke-Jian He}$^{b,}$$\footnote{\texttt{Corresponding author:kjhe94@163.com}}$ and
        \author{Qing-Quan Jiang}$^{a,}$$\footnote{\texttt{Corresponding author:qqjiangphys@yeah.net}}$
		\vskip 0.15in

$^{a}$\it{Physics and Astronomy College, China West Normal University, Nanchong 637000, People's Republic of China}\\
$^{b}$\it{Department of Mechanics, Chongqing Jiaotong University, Chongqing 400000, People's Republic of
China}\\
 	
	\end{center}
	\vskip 0.3in
	\begin{abstract}

In this paper, based on the action of a complex scalar field minimally coupled to a gravitational field, we numerically obtain a series of massive boson star solutions in a spherically symmetric background with a quartic-order self-interaction potential. Then, considering a thin accretion flow with a certain four-velocity, we further investigate the observable appearance of the boson star using the ray-tracing method and stereographic projection technique.
As a horizonless compact object, the boson star's thin disk images clearly exhibit multiple light rings and a dark central region, with up to five bright rings. As the observer's position changes, the light rings of some boson stars deform into a symmetrical ``horseshoe" or ``crescent" shape. When the emitted profile varies, the images may display distinct observational signatures of a ``Central Emission Region".
Meanwhile, it shows that the corresponding polarized images not only reveal the spacetime features of boson stars but also reflect the properties of the accretion disk and its magnetic field structure.
By comparing with black hole, we find that both the polarized signatures and thin disk images can effectively provide a possible basis for distinguishing boson stars from black holes. However, within the current resolution limits of the Event Horizon Telescope (EHT), boson stars may still closely mimic the appearance of black holes, making them challenging to distinguish at this stage.

\end{abstract}
	
	\thispagestyle{empty}
	\newpage
	\setcounter{page}{1}
	
\section{Introduction}\label{sec1}
Recently, the Event Horizon Telescope (EHT) collaboration has successfully released the 230 GHz images\cite{EventHorizonTelescope:2019dse,EventHorizonTelescope:2022wkp} and the polarized images of the central black holes in M$87$ and Sagittarius A\cite{EventHorizonTelescope:2021bee,EventHorizonTelescope:2024hpu}. These groundbreaking observations directly confirm the predictions of general relativity, precisely measure the black hole mass, and unveil the physical mechanisms governing the accretion disk's dynamics and jet formation, marking the dawn of a new era in compact object astrophysics. Ultra-compact objects, such as black holes, boson stars, and other theoretically predicted exotic dense stars, serve as crucial laboratories for probing extreme physics and testing gravitational theories\cite{Cunha:2022gde}.
In particular, the shadow and optical signatures of these compact objects not only reveal their spacetime characteristics but also encode critical information about magnetic field configurations, accreting matter distributions, and jet acceleration mechanisms. Thus, the deep exploration of the images of ultra-compact objects will not only allow for precise observation and distinction of compact objects in astrometry but may also provide key insights into probing extreme gravitational fields and understanding outstanding puzzles in fundamental physics.

Black holes, as typical ultra-compact objects, have shadow features (dark areas in EHT images) and photon rings (thin bright rings within luminous structures) that have been widely studied\cite{Wei:2013kza,Huang:2016qnl,Cunha:2017eoe,Perlick:2021aok}.
In general, the shadow of a spherically symmetric black hole is a circle, while rotating black holes exhibit ``D"-shaped features\cite{Perlick:2021aok}. The photon ring, composed of photons with multiple U-shaped turns, displays exponentially narrowed self-similar structures and may have a potential connection with the holographic principle\cite{Hadar:2022xag}.
Obviously, the image of a black hole depends on both the gravitational lensing effect and the characteristics of the light source. Based on celestial sources, one can observe the shadow and Einstein ring of a black hole\cite{Cunha:2015yba}.
In 2019, Wald et al. further studied the thin disk images of the Schwarzschild black hole and analyzed the properties of direct emission, lensed rings, and the photon ring\cite{Gralla:2019xty}.
This work was quickly extended to investigate optical and polarized images of various types of black holes in the framework of thick disks\cite{Huang:2024bar,Vincent:2022fwj}, jets\cite{Zhang:2024lsf,Papoutsis:2022kzp}, hotspot models\cite{Chen:2024ilc,Huang:2024wpj}, the general relativistic magnetohydrodynamic(GRMHD) simulation\cite{Wong:2022rqr}, and so on.
In addition, by using EHT observational data, black hole shadows can also be used to constrain the parameters of the relevant theories or models\cite{Chen:2019fsq,Guo:2025zca}.
Therefore, it is true that the shadow, photon ring, as well as the optical images of black holes and other compact objects, have received increasing interest both theoretically and experimentally in recent years\cite{Hu:2020usx,Narayan:2019imo,Himwich:2020msm,Zeng:2020dco,Gralla:2020yvo,Zeng:2020vsj,Peng:2020wun,Long:2020wqj,Qin:2020xzu,Peng:2021osd,Li:2021riw,
Okyay:2021nnh,Chael:2021rjo,Guerrero:2021ues,Bronzwaer:2021lzo,Li:2021ypw,Zeng:2021mok,Guo:2021bhr,Gan:2021pwu,Gan:2021xdl,Hou:2022eev,Meng:2023htc,Rosa:2023hfm,
He:2024qka,Li:2024ctu,Zeng:2022pvb,Guerrero:2022qkh,Chakhchi:2022fls,Guerrero:2022msp,Uniyal:2022vdu,DeMartino:2023ovj}.

However, considering the (arguably) unobservable nature of the event horizon¡ªthe most remarkable feature characterizing a black hole-and the theoretical and experimental uncertainties in observations of gravitational waves and black hole images, as well as inherent biases in data interpretation, the study of black hole ``mimickers" has emerged as a fast-growing area of research in recent years \cite{Cardoso:2019rvt}.
These mimickers, typically ultra-compact objects lacking an event horizon, can closely resemble black holes in many aspects and may even ``masquerade" as black holes under certain conditions. More importantly, the current EHT observations cannot rule out the possibility that the central object might be a horizonless compact star rather than a black hole, which makes the investigation of the shadows and optical signatures of such objects particularly significant. Among these compact objects, boson stars stand as a theoretically significant class of celestial bodies in modern physics. When the scalar field couples with the gravitational field, a stable localized soliton structure may form, which can remain stable in astrophysical environments and is always referred to as a boson star\cite{Liebling:2012fv}.
Boson star models were first explored in the 1960s, when Kaup obtained spherically symmetric solutions by coupling a complex scalar field to Einstein gravity, while Ruffini and Bonazzola studied a similar model with a real scalar field\cite{Kaup:1968zz,Ruffini:1969qy}.
Since then, various studies have been carried out on different aspects of boson stars, producing a series of significant scientific findings. For instance, a wide variety of boson star solutions have been constructed, including charged boson stars arising from coupling to the electromagnetic field\cite{Jetzer:1992tog}, Newtonian boson stars obtained in the weak-field limit\cite{Silveira:1995dh}, as well as rotating solutions endowed with angular momentum\cite{Li:2019mlk}.
Additionally, by considering a solitonic potential, the gravitational wave signals predicted from boson star binaries are studied in detail\cite{Bezares:2022obu}, with the results compared to those of black holes. Moreover, boson stars are also a possible candidate for dark matter, i.e., the dark matter halo can be produced by an N-body system of boson stars\cite{Liebling:2012fv,Pitz:2024xvh}.
In summary, boson stars play an important role in astrophysics and are a hot topic of current research.

In fact, since the release of black hole images by EHT, the images of boson stars, along with their observable features, have garnered increasing attention. Due to the absence of an event horizon, photons can pass through the interior of the star along geodesics, thereby naturally leading to images that differ noticeably from those of black holes. In 2022, following the idea of Wald, Rosa obtained the shadow images of spherically symmetric scalar bosons under a geometrically thin accretion disk\cite{Rosa:2022tfv}. They find that both the size of the shadow and the thin disk images are all heavily influenced by the emission
profile, and some boson stars may be able to produce black hole-like shadow images, even if they lack a photon ring\cite{Rosa:2022tfv}.
Then, based on the hot spot model, it has been found that an extra image can be produced in boson stars\cite{Rosa:2022toh}.
Additionally, by considering quartic and sixth-order self-interaction potentials, the images of boson stars orbited by a hot spot and a thin disk have been studied in \cite{Rosa:2023qcv}.
For an optically thick and geometrically thin disk, compact solitonic boson stars have the potential to mimic black holes in light of their shadow images\cite{Rosa:2024eva}.
However, it should be noted that in \cite{Rosa:2022tfv,Rosa:2023qcv}, when employing a thin disk as the light source, the static thin-disk accretion model is typically adopted. Recently, we studied the images of solitonic and mini boson stars by considering the thin disk\cite{Zeng:2025xoe,He:2025qmq}, here the accretion flow moves along timelike circular orbits. At present, within the framework of a dynamic thin disk, it is obvious that the image of boson stars with the quartic-order self-interaction potential remains unknown. On the other hand, polarized images have been widely studied in various kinds of black holes, as they can help us gain deeper insights into the astrophysical and geometric properties of black hole accretion flows by comparing theoretical polarized patterns with observed results\cite{Lupsasca:2018tpp,Himwich:2020msm}.
In 2021, using the Penrose-Walker constant, the polarized images of axisymmetric fluid were investigated for a Kerr black hole with a toy model\cite{Gelles:2021kti}.
Many works have been carried out on the polarization images of different types of black holes and horizonless compact objects, see Refs. \cite{Qin:2021xvx,Hu:2022sej,Liu:2022ruc,Qin:2022kaf,Delijski:2022jjj,Deliyski:2023gik,Zhang:2023cuw,Qin:2023nog,Shi:2024bpm} and the references therein. Especially for horizonless compact objects, such as traversable wormholes and naked singularities, it has been shown that observable polarized images can be used to distinguish them from black holes\cite{Delijski:2022jjj,Deliyski:2023gik}.
Even so, the polarized images of boson stars under thin disk conditions remain unclear, and it is still unknown whether their features can distinguish them from black holes.
Combined with the above fact, the motivation of this paper is to study the observable features of boson stars with a quartic-order self-interaction potential. By exploring their accretion disk images and polarization features, we aim to determine whether these boson stars can mimic black holes under current resolution limits, and whether future high-precision astronomical observations can effectively distinguish between boson stars and black holes.

The paper is organized as follows: Section \ref{sec2} introduces the action with a complex scalar field and then derives the numerical solution of the massive boson stars. Section \ref{sec3} details ray-tracing method and focuses on images under the thin disk. Section \ref{sec4} is devoted to investigating the characteristics of polarized images of boson stars. Section \ref{sec5} provides a comparison with the Schwarzschild black hole. Finally, Section \ref{sec6} concludes with a summary.

\section{The solutions and fitting of boson stars}\label{sec2}
\begin{figure}[!h]
\makeatletter
\renewcommand{\@thesubfigure}{\hskip\subfiglabelskip}
\makeatother
\centering 
\vspace{-0.2cm}
\subfigure[$ $]{
\setcounter{subfigure}{0}
\subfigure[$(a)$]{\includegraphics[width=.3\textwidth]{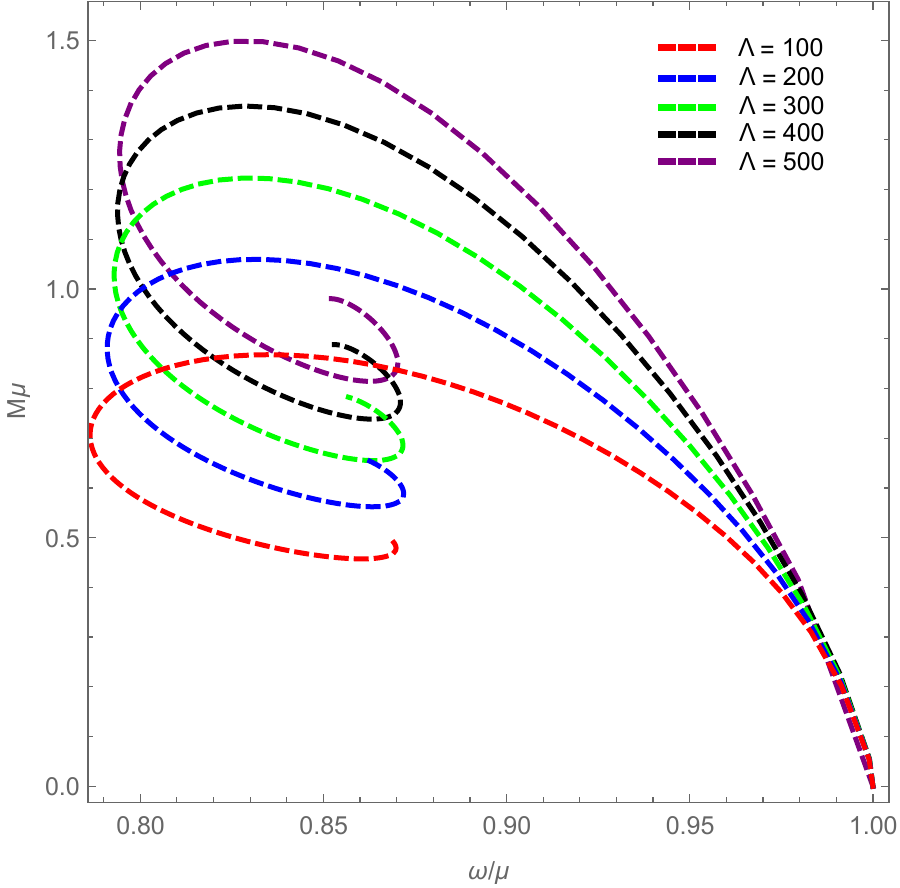}}
\subfigure[$(b)$]{\includegraphics[width=.297\textwidth]{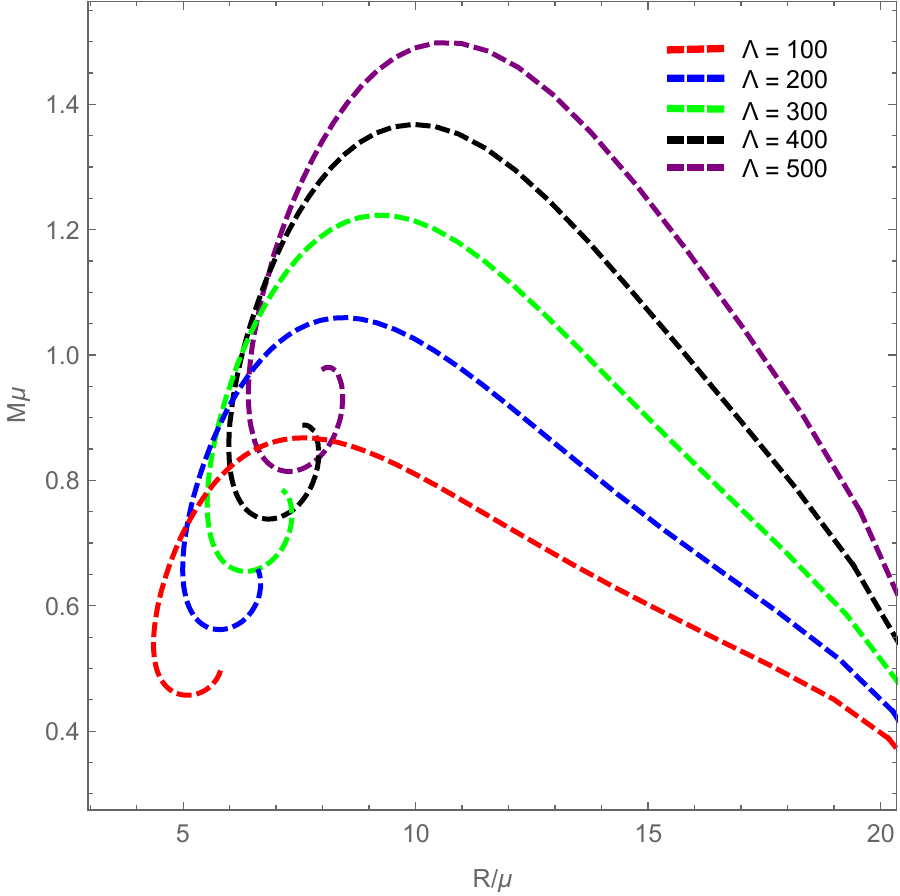}}
\subfigure[$(c)$]{\includegraphics[width=.305\textwidth]{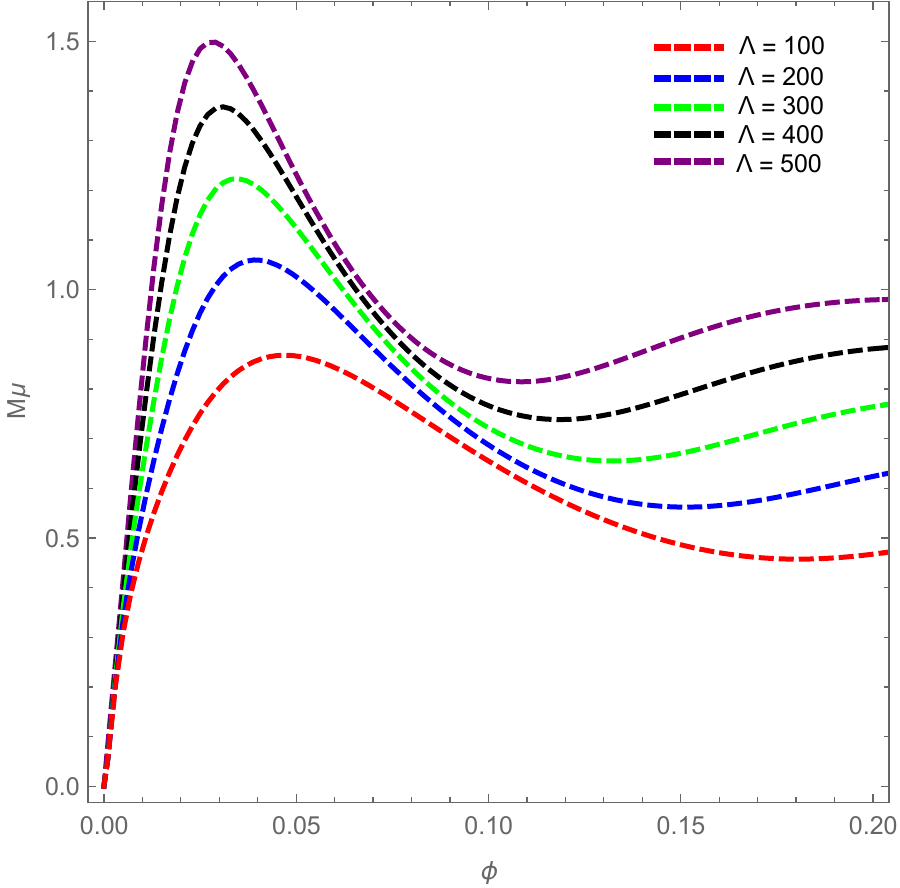}}
}
\vspace{-0.4cm}
\caption{\label{fig1} The mass versus the frequency $\omega$, the radius $R$, and the  scalar field $\phi$.}
\end{figure}

In this section, we will first study scalar boson stars using the quartic-order self-interaction potential \cite{Colpi:1986ye}, commonly referred to as massive boson stars. For a self-gravitating solution of a massive and complex scalar field $\Phi$, it can be constructed within the framework of the Einstein-Klein Gordon theory. In general, this theory is described by the following action $S$,
\begin{align}\label{e1}
S = \int d^4 x \sqrt{-g} \left(\frac{R}{2 \kappa}-\nabla_a \Phi^* \nabla^a \Phi - V (\left| \Phi \right|^2)  \right).
\end{align}
In Eq.(\ref{e1}), the coupling constant $\kappa$ is typically chosen as $\kappa = 8\pi$, and $g_{\mu\nu}$ is the spacetime metric, with its determinant denoted by $g = \det(g_{\mu\nu})$. And, $\Phi^*$ is the complex conjugate of the scalar field $\Phi$, while $R$ and $V$ are the scalar curvature and scalar potential, respectively. For convenience, the geometrized units $G = c = 1$ are used throughout the paper, where $G$ is the gravitational constant and $c$ denotes the speed of light. By taking a variation of the action (\ref{e1}) with respect to the metric $g_{\mu\nu}$ and the scalar field $\Phi$, it is easy to obtain the following equations of motion, which are,
\begin{align}\label{e2}
R_{ab} - \frac{1}{2} g_{ab} R = \kappa T_{ab}, \quad
\nabla_a \nabla^a \Phi = \Phi \frac{d V}{d|\Phi|^2},
\end{align}
where $R_{ab}$ represents the Ricci tensor, and $T_{ab}$ is the energy-momentum tensor of the scalar field $\Phi$. The form of  $T_{ab}$ is
\begin{align}\label{e3}
T_{ab} = \nabla_a \Phi^* \nabla_b \Phi + \nabla_b \Phi^* \nabla_a \Phi - g_{ab}\left(\nabla_c \Phi^*  \nabla^c \Phi +V  \right).
\end{align}
Since we focus on such a general static and spherically symmetric solution, the following ansatz for the metric $g_{\mu \nu}$ is considered,
\begin{align}\label{e4}
ds^2=-A(r)dt^2 +B(r)^{-1}dr^2 + r^2(d\theta^2 + \sin^2{\theta} d\varphi^2),
\end{align}
where $A(r)$, $B(r)$ are all the function of $r$.
And, we carry out the scalar field $\Phi$ as,
\begin{align}\label{e5}
\Phi(r,t) = \phi(t) e^{i \omega t}.
\end{align}
By substituting Eqs.(\ref{e4}) and (\ref{e5}) into the field equations (\ref{e2}), the equations of motion for the system can be obtained. Their form is as follows,
\begin{align}\label{e6}
&A'(r) = \frac{A(r)[1-B(r)]}{r B(r)} + 8\pi \left[\frac{\omega^2 \phi^2}{B(r)} +A(r)\phi'^2 - V \frac{A(r)}{B(r) }  \right], \nonumber \\
&B'(r)=\frac{1-B(r)}{r} -8\pi \left[\frac{\omega^2 \phi^2}{A(r)} +B(r) \phi'^2 +V  \right], \\
&\phi'' + \left[\frac{2}{r} + \frac{1}{2}\left[ \frac{A'(r)}{A(r)}+ \frac{B'(r)}{B(r)}  \right]  \right]\phi' + \frac{1}{B(r)}\left[\frac{\omega^2}{A(r)} - \frac{dV}{d\phi^2}   \right]\phi=0.\nonumber
\end{align}
For the boson star, we can characterize the solution based on their total mass $M$ and radius $R$. Both two quantities can be defined through the mass function $\mathbf{M}(r)$, which is given by $\mathbf{M}(r) = \frac{r}{2}\left[1- B(r)\right]$. The ADM total mass is defined as $M=\mathbf{M}(r\rightarrow \infty)$. Generally, we consider the radius $R$ of boson star to be located at $R=0.98M$. Since the mass of the boson star is always relatively concentrated, the calculation results for the radius of this star, defined by different methods, will not show significant differences. In this paper, our focus is mainly on massive boson stars, where the potential function $V$ is chosen as
\begin{align}\label{e7}
V = \mu^2 \phi^2 + \Lambda \phi^4,
\end{align}
with $\mu$ is a constant which corresponds to the mass of the scalar field $\Phi$, and the coupling constant $\Lambda$ denotes the intensity of the quartic self-interaction of $\Phi$. Because the analytical solution becomes a very difficult task in Eq.(\ref{e6}), we employ the simplest shooting method to find a numerical solution. In our unit system, the only dimensional quantity is the mass $[M]$, and the equations of motion remain invariant under the following transformation,
\begin{align}\label{e8}
r \rightarrow \mu r, \quad B \rightarrow B, \quad A \rightarrow \mu^2 A, \quad  V \rightarrow \mu^{-2} V.
\end{align}
For simplicity, we set $\mu=1$. For the asymptotic behavior of the solution, we require it to exhibit asymptotic flatness similar to that of Schwarzschild spacetime at infinity, i.e.,
\begin{align}\label{e9}
A(r \rightarrow \infty) \sim A_{\infty} \left(1-\frac{2M}{r}  \right), \quad B(r \rightarrow \infty) \sim 1-\frac{2M}{r}, \quad \phi(r \rightarrow \infty) =0.
\end{align}
Considering the asymptotic behavior at $r=0$, we expect this solution to remain non-divergent. At this point, by expanding $A(r)$, $B(r)$ and $\phi$ into power series of $r$, it yields
\begin{align}\label{e10}
A(r \rightarrow 0) = A_0, \quad B(r \rightarrow 0) = 1,  \quad\phi(r \rightarrow 0) = \phi_0,
\end{align}
where, $A_0$ and $A_{\infty}$ is not independent. In addition, since the equations are independent of time $t$, we can reparametrize $t$, $A(r)$, and $\omega$ by choosing an arbitrary constant. Under the suitable transformation, the entire system remains unchanged, so we can simply set the initial boundary condition $A_0 = 1$ as one numerically solves the equation of motion.
Based on the aforementioned boundary conditions, we use the shooting method to numerically solve the equation of motion (\ref{e6}). By choosing different values of $\Lambda$ and $\phi_0$, ten specific boson stars are considered. As $\Lambda=200$, the values $\phi_0 = (0.06, 0.09, 0.12, 0.15, 0.18)$ correspond to the boson stars (PBS1, PBS2, PBS3, PBS4, PBS5) respectively. For the case of $\phi_0=0.15$ with the choice of $\Lambda = (100, 200, 300, 400, 500)$, the related boson stars are labeled as ($\Lambda$BS6, $\Lambda$BS7, $\Lambda$BS8, $\Lambda$BS9, $\Lambda$BS10).
In Figure \ref{fig1}, we employ the numerical solution to present the relationship between $M$ and ($\omega$, $R$, $\phi$) for different values of $\Lambda$.

\vspace{-0.1cm}

It can be seen that with the decrease of the parameters $\omega$ and $R$, the mass of the boson stars initially increases, but then decreases and exhibits a spiral behavior. Meanwhile, when the scalar field $\phi$ increases, the mass first grows, then decays, and afterwards grows again, eventually stabilizing at a constant value. Additionally, it can be observed that the parameter $\Lambda$ decreases the mass of the boson star for the fixed values of $\omega$, $R$, and $\phi$.

\begin{figure}[!h]
\makeatletter
\renewcommand{\@thesubfigure}{\hskip\subfiglabelskip}
\makeatother
\centering 
\subfigure[$ $]{
\setcounter{subfigure}{0}
\subfigure[$$]{\includegraphics[width=.4\textwidth]{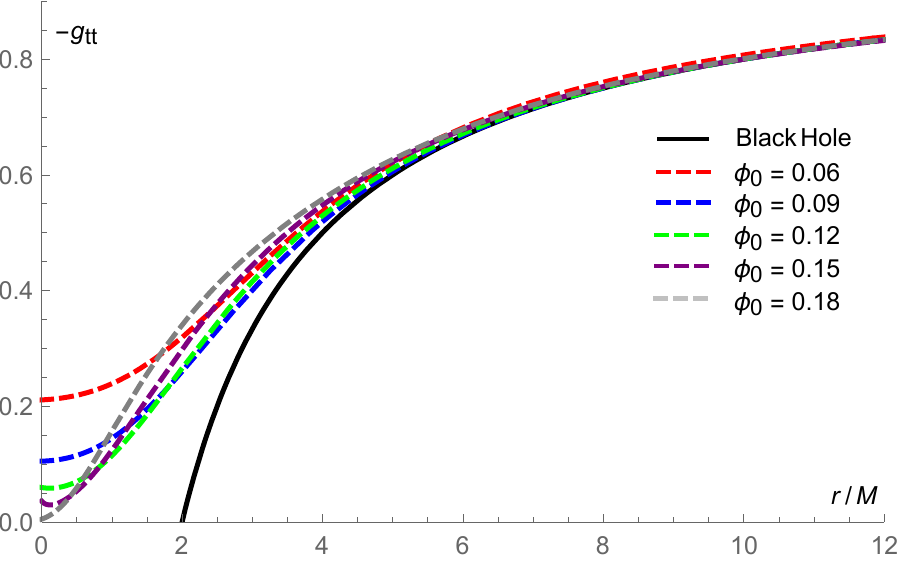}} \qquad
\subfigure[$$]{\includegraphics[width=.4\textwidth]{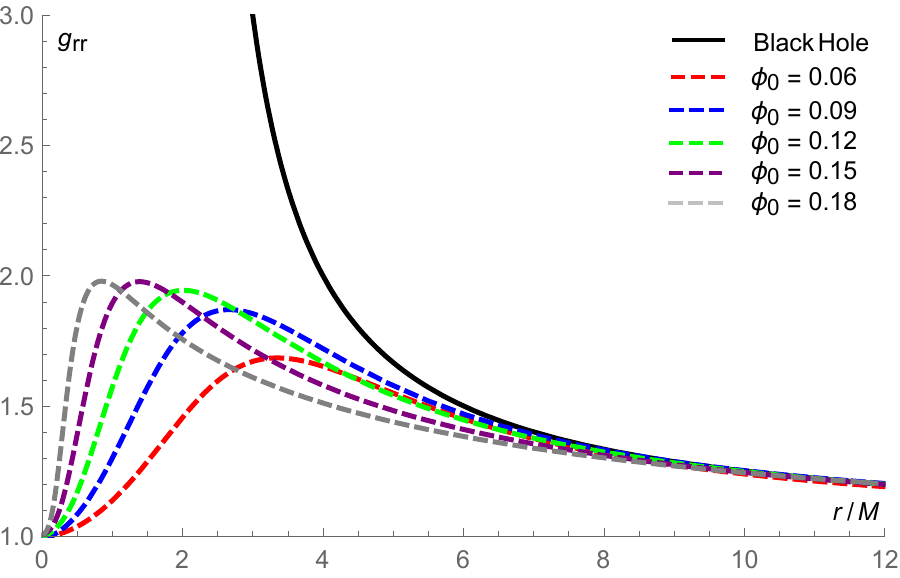}}}
\vspace{-0.4cm}
\subfigure[$ $]{
\setcounter{subfigure}{0}
\subfigure[$(a)$: The metric function $g_{tt}$.]{\includegraphics[width=.4\textwidth]{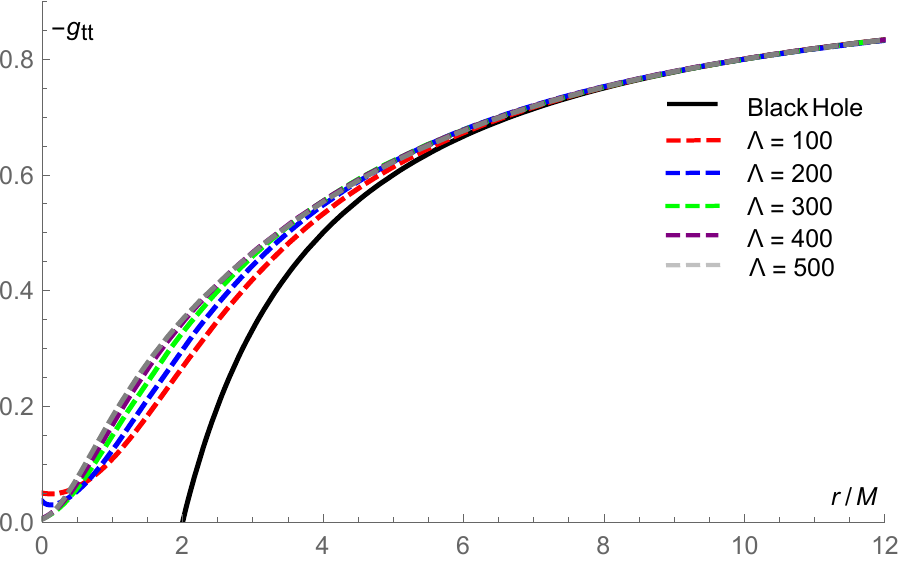}} \qquad
\subfigure[$(b)$: The metric function $g_{rr}$.]{\includegraphics[width=.4\textwidth]{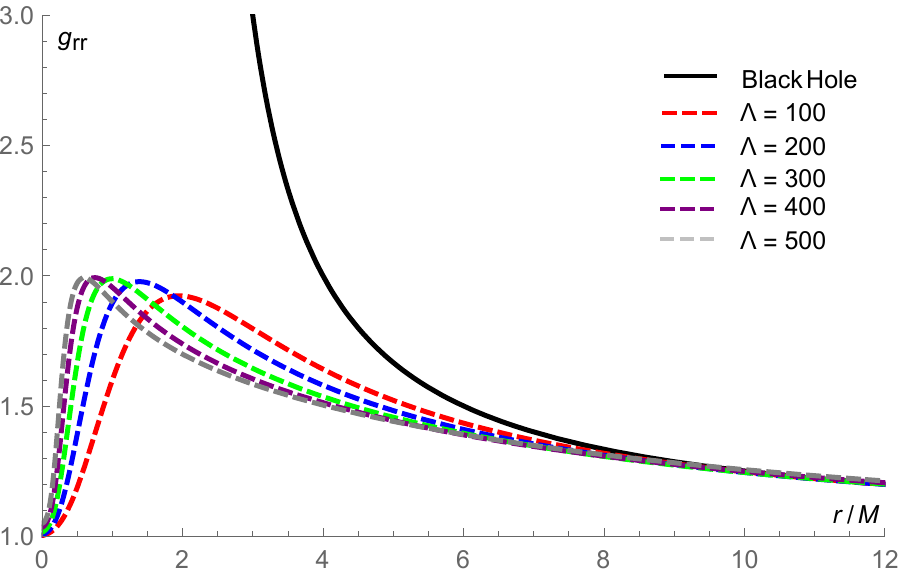}}}
\vspace{-0.2cm}
\caption{\label{fig4} The metric functions $g_{tt}$ and $g_{rr}$. }
\end{figure}

For numerical solutions, it is not very convenient to directly use the numerical metric for imaging those boson stars under the thin disk. Therefore, we adopt a fitting approach to obtain an analytical metric, and the fitting form is,
\begin{align}\label{e11}
&g_{tt} = -\exp{\left\{b_7   \left[\exp{\left( -\frac{1 + b_1 r +b_2 r^2}{b_3 + b_4 r +b_5 r^2 + b_6 r^3} \right)}  -1    \right]     \right\}},\\
&g_{rr} = \exp{\left\{a_7   \left[\exp{\left( -\frac{1 + a_1 r +a_2 r^2}{a_3 + a_4 r +a_5 r^2 + a_6 r^3} \right)}  -1    \right]     \right\}}.
\end{align}
Since those coefficients need to satisfy the boundary conditions of the equations of motion, they are not completely independent, which naturally imposes corresponding constraints between the coefficients. Those coefficients have been presented in Tables 1 and 2, and the fitting of the metric functions $g_{tt}$ and $g_{rr}$ for different values of $\phi_0$ and $\Lambda$ have been shown in Figure \ref{fig4}.

\begin{center}\label{T1}
{\footnotesize{\bf Table 1.} The Boson stars for $\Lambda = 200 $.\\
\vspace{1mm}
\begin{tabular}{ccccccccccc}
\hline \hline &{$b_1$} &{$b_2$} &{$b_3$} &{$b_4$} &{$b_5$} &{$b_6$} &{$b_7$} &{$M$} &{$R$} \\ \hline
{PBS1}  &{-0.00554} &{0.05020} &{-0.28269} &{0.01424} &{-0.01814}  &{-0.00122}  &{-0.04662} &{0.95883}  &{6.32394}  \\
{PBS2}  &{0.14727} &{0.18049}  &{-0.35860}  &{-0.05541}  &{-0.09006}  &{-0.01794}  &{-0.14769}   &{0.74295}   &{5.12562}  \\
{PBS3}  &{0.70538}  &{0.13360}  &{-1.8162}  &{-1.00257}  &{-2.12412}   &{-0.42257}  &{3.84165}  &{0.60730} &{5.08224} \\
{PBS4}  &{8.49375}  &{4.94370}  &{-0.40134}  &{-3.08499}   &{-3.68206}   &{1.31779}  &{-0.29968} &{0.56214} &{5.74655} \\
{PBS5}  &{-0.41599} &{0.18015} &{-0.37639}  &{-0.19806}  &{0.09174}  &{-0.06143}  &{-0.40217}  &{0.58970} &{6.46439}   \\
\hline
&{$a_1$} &{$a_2$} &{$a_3$} &{$a_4$} &{$a_5$} &{$a_6$} &{$a_7$} &{$M$}  &{$R$}\\ \hline
{PBS1}   &{-2.58625}  &{-54.8059}  &{-748.583} &{-383.377} &{208.025} &{-73.3757} &{-2.32830} &{0.95883}  &{6.32394} \\
{PBS2}   &{-2.47113}  &{-100.713}  &{-124.652} &{-138.843} &{136.856} &{-67.3973} &{-0.97188} &{0.74295}   &{5.12562} \\
{PBS3}   &{-41.5062}  &{-116.124}  &{7.89008}  &{7.28212}  &{33.5177} &{4.52372}  &{0.03590}  &{0.60730}   &{5.08224} \\
{PBS4}   &{21.0478}   &{50.2354}   &{-0.68063} &{-2.27999} &{-8.49571} &{-0.32132} &{0.00143} &{0.56214}  &{5.74655} \\
{PBS5}   &{43.0469}   &{278.352}   &{-1.00565} &{-4.45958} &{-49.1396}  &{-1.81578} &{0.00197} &{0.58970}  &{6.46439}   \\

\hline \hline
\end{tabular}}
\end{center}

\begin{center}\label{T2}
{\footnotesize{\bf Table 2.} The Boson stars for $\phi_0 = 0.15 $.\\
\vspace{1mm}
\begin{tabular}{ccccccccccc}
\hline \hline &{$b_1$} &{$b_2$} &{$b_3$} &{$b_4$} &{$b_6$} &{$b_6$} &{$b_7$} &{$M$} &{$R$} \\ \hline
{$\Lambda$BS6} &{1.01575}   &{0.12013}   &{-1.21450}  &{-0.95280}  &{-2.43138}  &{-0.28945}  &{-2.34367}  &{0.48636}  &{4.49651}  \\
{$\Lambda$BS7} &{8.49375}   &{4.94370}   &{-0.40134}  &{-3.08499}  &{3.68206}  &{-1.31778}  &{-0.29968}  &{0.56214}  &{5.74655}    \\
{$\Lambda$BS8} &{-0.02109}  &{0.02753}   &{-0.50633}  &{-0.48946}  &{0.00257}  &{-0.01707}  &{-0.83151}   &{0.67042} &{6.88161}   \\
{$\Lambda$BS9} &{-0.28493}  &{0.10297}   &{-0.40875}  &{-0.19921}  &{0.06179}  &{-0.03157}  &{-0.48359}   &{0.78857}  &{7.74236}   \\
{$\Lambda$BS10} &{-0.16015} &{0.08183}  &{-0.58309}  &{-0.55194}  &{0.10988}  &{-0.05345}  &{-1.17928}   &{0.90262} &{8.39121} \\
\hline
&{$a_1$} &{$a_2$} &{$a_3$} &{$a_4$} &{$a_5$} &{$a_6$} &{$a_7$} &{$M$}  &{$R$}\\ \hline
{$\Lambda$BS6} &{63.6905}  &{53.1529}   &{-2.28649}  &{-7.97932}  &{-11.3295}  &{-0.51836}  &{0.00318}   &{0.48636}  &{4.49651} \\
{$\Lambda$BS7} &{21.0478}  &{50.2354}  &{-0.68063}  &{-2.27999}  &{-8.49571}  &{-0.32132}  &{0.00143}   &{0.56214}  &{5.74655}   \\
{$\Lambda$BS8} &{4.61062}  &{46.8521}   &{-0.32483}  &{-0.21878}  &{-6.96248}  &{-0.19380}  &{0.00071}   &{0.67042}   &{6.88161} \\
{$\Lambda$BS9} &{-0.26133} &{42.9810}  &{-0.238929}  &{0.328775}  &{-6.09154}  &{-0.12839}  &{0.000509}  &{0.78857}  &{7.74236}  \\
{$\Lambda$BS10} &{-2.27039}&{43.2295}  &{-0.18505}  &{0.48304}  &{-5.47852}  &{-0.08328}  &{0.00022}   &{0.90262}   &{8.39121} \\
\hline \hline
\end{tabular}}
\end{center}

\vspace{-0.1cm}

{Form the Tables 1 and 2, it is true that the mass of boson stars always increase with the parameter $\Lambda$, while decrease in the region of $\phi_0 \in (0.0398,0.15)$, and increase in ($0.15,0.25$), which are coincide with that given in Figure \ref{fig1}. }
In Figure \ref{fig4}, the black lines  represent the corresponding  $g_{tt}$ and $g_{rr}$ of Schwarzchild black hole.
It is obvious that the metric function $g_{tt}$ and $g_{rr}$ are always closer and closer to the case of Schwarzschild black hole when $r$ is bigger and bigger.
And, the peak of function $g_{rr}$ increase with $\Lambda$, while decrease with $\phi_0$.
Also, the functions $g_{tt}$ seem to have a smaller value at the point $r=0$ for a larger $\phi_0$ or $\Lambda$.

\section{Images of boson stars under the thin disk }\label{sec3}
In this section, we will continue to investigate the observable appearance of boson stars in the presence of a accretion disk that is both optical and geometric thin under different emission models. The thin disk is located at the equatorial plane, and the plasma on the disk moves in circular orbits along timelike trajectories. For this type of particle, we have
$g_{\mu\nu}\dot{x}^{\mu}\dot{x}^{\nu}=-m^2$, where $m=1$ represents the mass of particle. Additionally, since the spacetime of the boson star is a spherically symmetric static spacetime, we can relatively easily derive the corresponding effective potential, which is given by
\begin{align}
\Dot{r}^2 = V(r), \quad V(r) = -m^2 -g_{rr}^{-1} \left(\frac{E^2}{ g_{tt}} + \frac{ L^2}{r^2}\right),
\end{align}
as well as
\begin{align}
    E= -\frac{\partial \mathcal{L}}{\partial \Dot{t} }  =-g_{tt} \Dot{t}, \quad L=\frac{\partial \mathcal{L}}{\partial \Dot{\varphi} }  =r^2 \Dot{\varphi},
\end{align}
where, $\mathcal{L} = \cfrac{1}{2} g_{\mu\nu}\dot{x}^{\mu}\dot{x}^{\nu}$ is the Lagrangian, $E$ and $L$ correspond to the energy and angular momentum of the particle.
And, $\dot{x}^{\mu}$ the derivative of spacetime coordinates with respect to the affine parameter $\lambda$.
In general, the innermost stable circular orbit (ISCO) of boson stars can be determined by $V(r)=0\mid_{r=r_{ISCO}}$, $\partial_rV(r)=0\mid_{r=r_{ISCO}}$ and $\partial^2_rV(r)=0\mid_{r=r_{ISCO}}.$
At positions well inside the ISCO, we assume that there is no presence of accreting material, meaning there is no light source.
Near and outside the ISCO, the accretion flow moves along timelike circular orbits\footnote{In this case, we regard the accretion flow moves counterclockwise.}, it's four-velocity is
 \begin{align}
    u^{\mu}= \left(-\sqrt{\frac{2 g_{tt}^2}{r \partial_rg_{tt} -2 g_{tt}}},0,0,\sqrt{\frac{r^3 \partial_rg_{tt} }{2 g_{tt}- r\partial_rg_{tt} }}\right)
\end{align}
On the other hand, it is well known that the motion of photons is described by the geodesic equation. In this paper, we numerically solve the geodesic equation with its Hamiltonian canonical form. When a photon is emitted from a light source and reaches an observer, a locally static coordinate system is typically used to describe the four-momentum of photon. This coordinate system established at the observer's position is commonly referred to as the zero angular momentum observer (ZAMO) tetrad. For the boson star spacetime in this paper, its form is as follows,
 \begin{align}
   & e_{(t)}=\left(\sqrt{-\frac{1}{g_{tt}}},0,0,0\right),  \quad e_{(r)}= \left(-\sqrt{\frac{1}{g_{rr}}},0,0,0\right),  \\
  & e_{(\theta)}= \left(0,\sqrt{\frac{1}{g_{\theta \theta}}},0,0\right), \quad e_{(\varphi)}= \left(-\sqrt{\frac{1}{g_{\varphi\varphi}}},0,0,0\right).
\end{align}
The four-momentum of photons can be reexpressed as $p_{(\mu)} = p_{\nu} e^{\nu}_{(\mu)}$ in this ZAMO tetrad, where $ p_{(\mu)}$ and $p_{\nu}$ denote the four-momentum in ZAMO and {Boyer-Lindquist coordinate} systems, respectively.
Meanwhile, we will also construct a celestial coordinate system ($\Theta,\Phi$) at the observer's position by using the magnitude of the photon's momentum as the radius, so that the position of the photon on the screen can be determined with stereographic projection techniques.
According to \cite{Hu:2020usx}, the celestial coordinates can be expressed as
\begin{align}
    \cos \Theta= \frac{p^{(r)}}{p^{(t)}}, \quad \tan \Phi=\frac{p^{(\varphi)}}{p^{(\theta)}}.
\end{align}
On the screen, the coordinates of Cartesian coordinates system are
\begin{align}
    X= - 2 \tan \frac{\Theta}{2} \sin \Phi, \quad Y= - 2 \tan \frac{\Theta}{2} \cos \Phi.
\end{align}
Therefore, if the pixel points on the screen are determined and the field of view ($\alpha_{fov}$) at the observer's position is given, we can use backward tracing techniques to trace back to the position of the light source\cite{Hu:2020usx}.

\begin{figure}[!h]
\makeatletter
\renewcommand{\@thesubfigure}{\hskip\subfiglabelskip}
\makeatother
\centering 
\vspace{-0.2cm}
\subfigure[$ $]{
\setcounter{subfigure}{0}
\subfigure[]{\includegraphics[width=.4\textwidth]{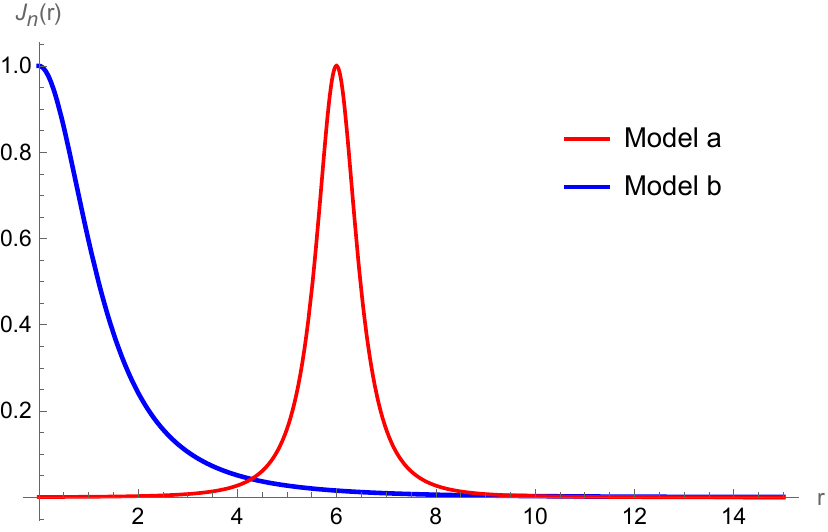}}}
\vspace{-0.6cm}
\caption{\label{figbpm} Intensity profiles for Model a and b.}
\end{figure}

To study the image of boson stars, we need to find an appropriate source model. Currently, based on astronomical observations, a geometrically and optically thin accretion disk is commonly used as the light source to investigate the image features of black holes. In 2019, by considering the presence of thin accretion disk in the equatorial plane of black hole, Wald found that the black hole image consists of a direct image, a lensing ring, and a photon ring\cite{Gralla:2019xty}. Based on this concept, more and more researches have been studied subsequently in various kinds of matter fields and gravitational theories. Since both black holes and boson stars are compact objects, this paper can also adopts a geometrically and optically thin accretion disk as the source model to study the image characteristics of boson star spacetimes.
If a light ray intersects the accretion disk $n$ times, and each intersection will gain additional luminosity. So, as we ignored the reflection effects and the thickness of the accretion disk, then the total observed light intensity on the screen will be a summation, that is,
\begin{align}
   I_{v_o}= \sum_{n=1}^{N_{max}} g_n^3 J_n,
\end{align}
with $v_o$ is the observed frequency, $N_{max}$ is the maximum number of intersections and $J_n$ is the emissivity. Here, we set $J_n$ to be the unbounded Johnson Distribution\cite{Gralla:2020srx}, which is
\begin{align}\label{jn}
   J_n= \frac{e^{-\frac{1}{2} \left( \gamma + \operatorname{arcsinh} \left[ \frac{r-\beta}{\sigma}\right]  \right)^2 }}{\sqrt{(r - \beta)^2 + \sigma^2}},
\end{align}
with $\gamma$, $\beta$ and $\sigma$ are related to the accretion disk model, i.e., $\gamma$ represents the rate of growth of the intensity from infinity to its peak, $\beta$ is the position of the peak of intensity, as well as $\sigma$ controls the dilation of the profile. This choice (\ref{jn}) was originally developed to match the results of GRMHD for Kerr black holes.
In addition, the redshift factor $g_n$ can be expressed as
\begin{align}\label{gn}
   g_n= -\frac{1}{u^{\mu} g^{\mu\nu} p_{\mu}},
\end{align}
where, the value of $g^{\mu\nu}$ is given at position $(0,r,\pi/2,0)$. Based on the camera model and the pixel settings on the screen\cite{Hu:2020usx}, and given the field of view of the camera $\alpha_{fov}$, we first numerically solve the photon motion equation using ray-tracing techniques.
Next, with the aid of the accretion disk emission model (\ref{jn}) and the redshift factor (\ref{gn}), we can obtain the image of the boson star for the counterclockwise accretion flow as observed by the observer at position $(0,200,\pi/2,0)$.
In this paper, we employ two different accretion disk emission models, which are\textbf{ Model a}: $\gamma=0$, $\beta=6M$ and $\sigma=M$;\textbf{ Model b}: $\gamma=0$, $\beta=0$ and $\sigma=1.25M$. For these two models, we present them intuitively in Figure \ref{figbpm}.

\vspace{-0.1cm}

\begin{figure}[htp]
\makeatletter
\renewcommand{\@thesubfigure}{\hskip\subfiglabelskip}
\makeatother
\centering 
\subfigure[]{
\setcounter{subfigure}{0}
\subfigure[$\Lambda=200,\phi_0=0.06$]{\includegraphics[width=.24\textwidth]{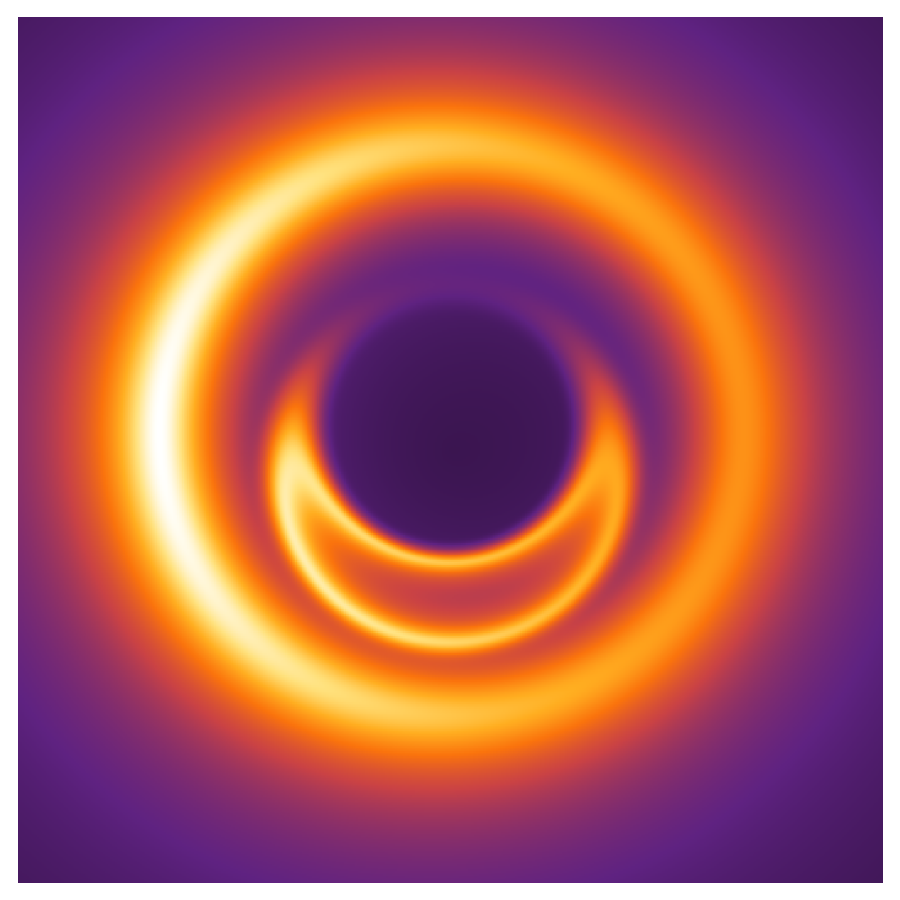}}
\subfigure[$\Lambda=200,\phi_0=0.09$]{\includegraphics[width=.24\textwidth]{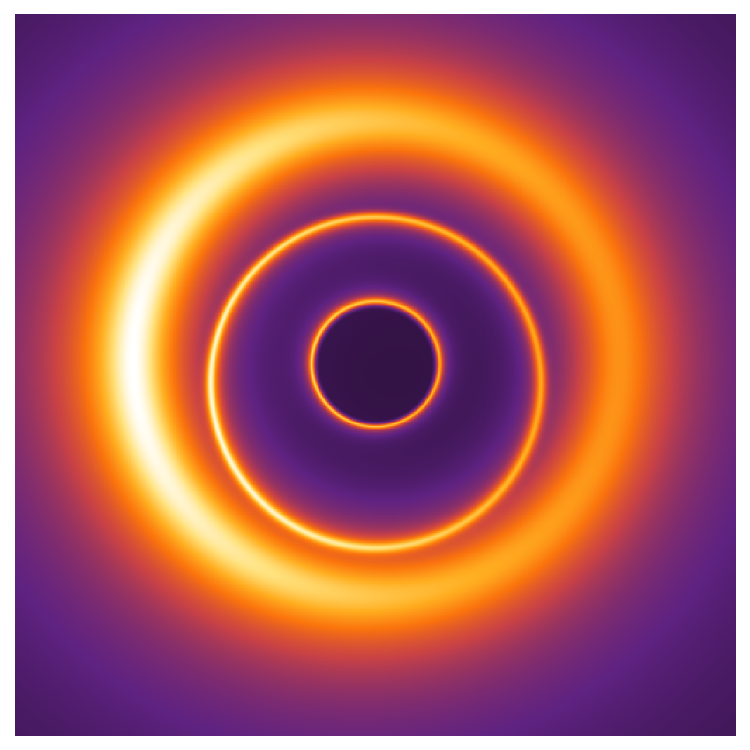}}
\subfigure[$\Lambda=200,\phi_0=0.15$]{\includegraphics[width=.24\textwidth]{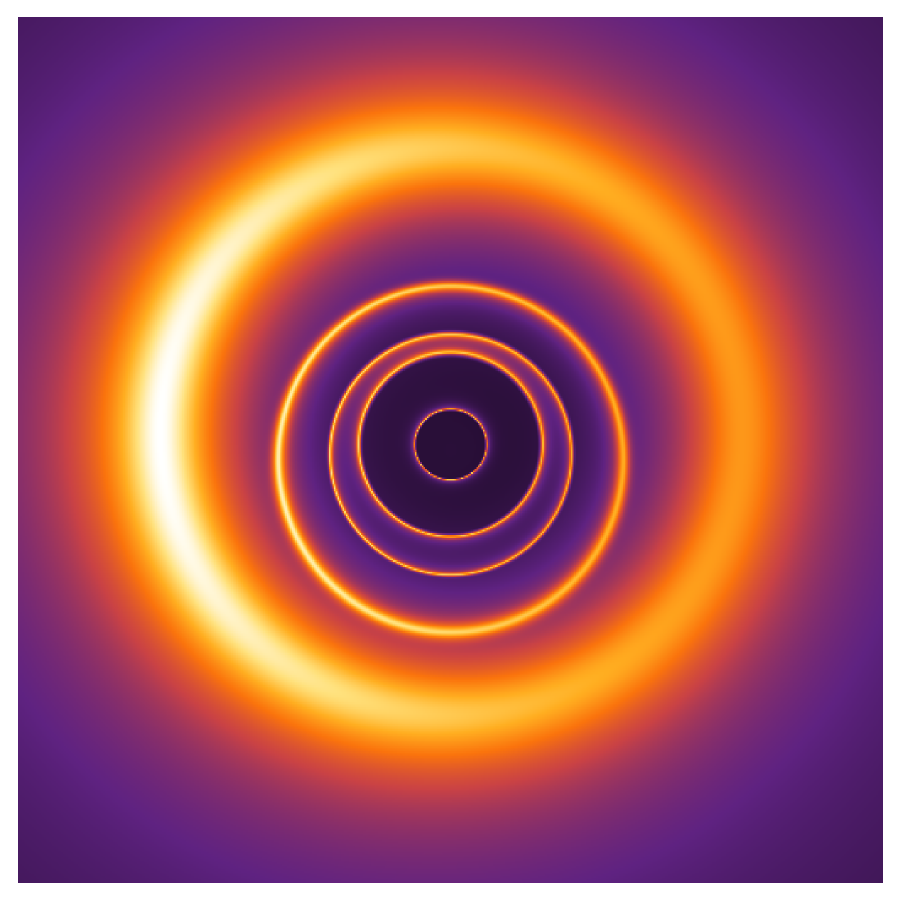}}
\subfigure[$\Lambda=200,\phi_0=0.18$]{\includegraphics[width=.24\textwidth]{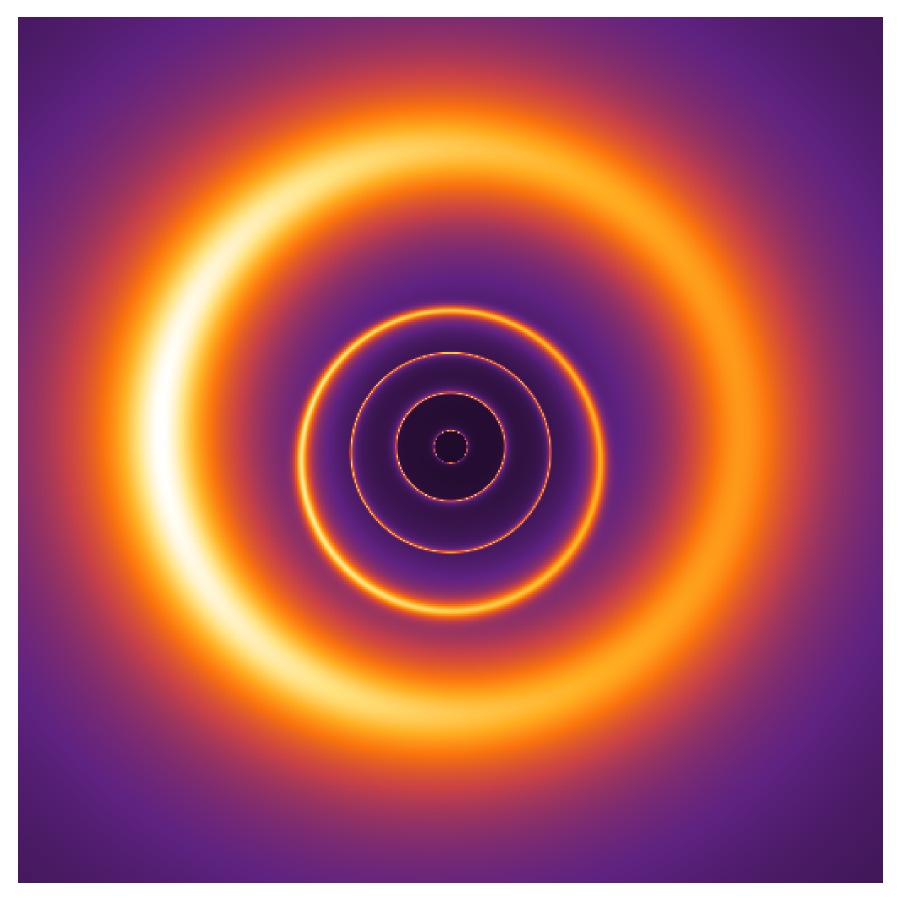}}
}
\subfigure[]{
\setcounter{subfigure}{0}
\subfigure[$\Lambda=100,\phi_0=0.15$]
{\includegraphics[width=.24\textwidth]{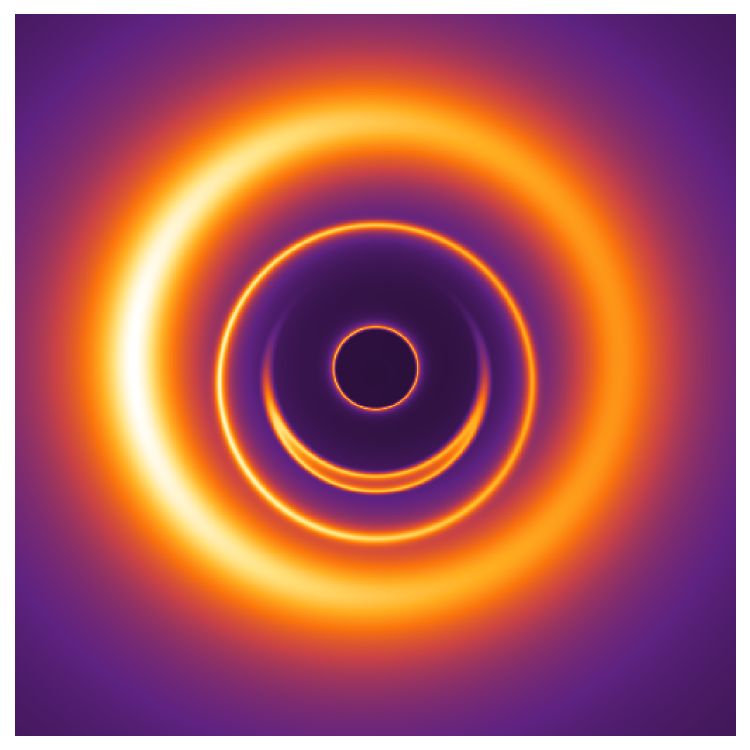}}
\subfigure[$\Lambda=200,\phi_0=0.15$]
{\includegraphics[width=.24\textwidth]{200015-17d.pdf}}
\subfigure[$\Lambda=300,\phi_0=0.15$]{\includegraphics[width=.24\textwidth]{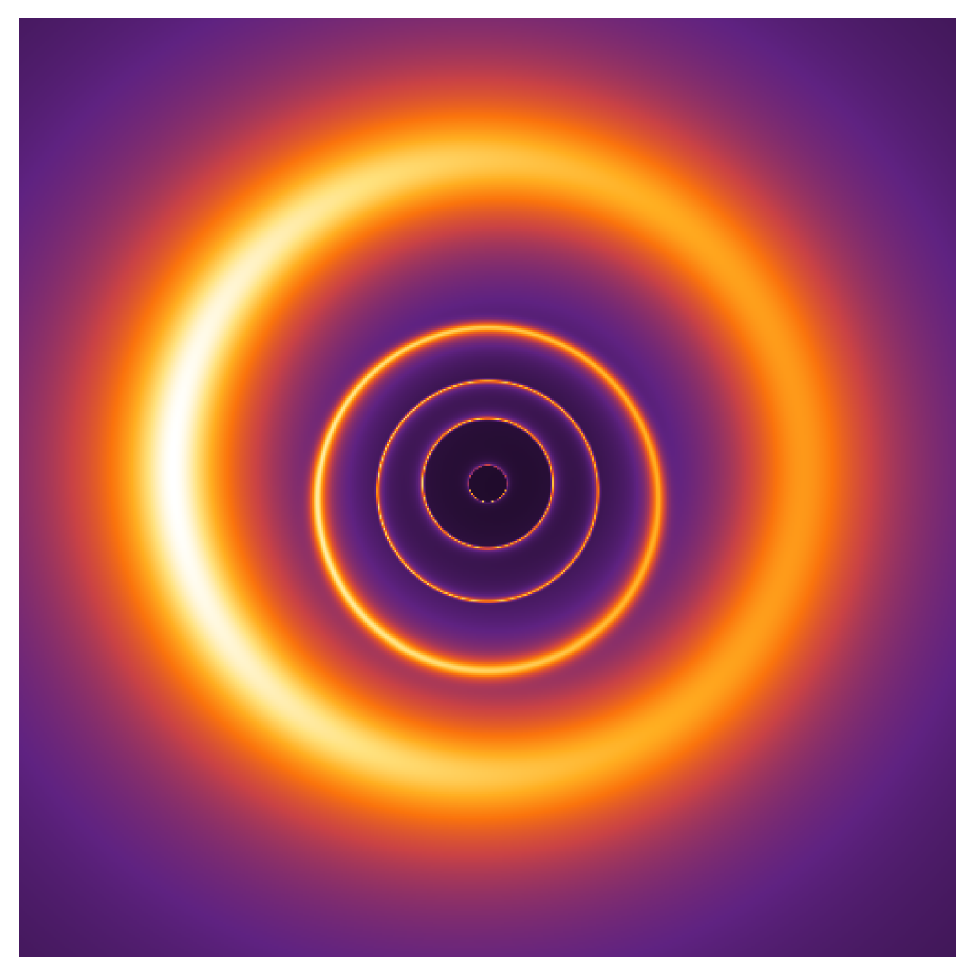}}
\subfigure[$\Lambda=500,\phi_0=0.15$]{\includegraphics[width=.24\textwidth]{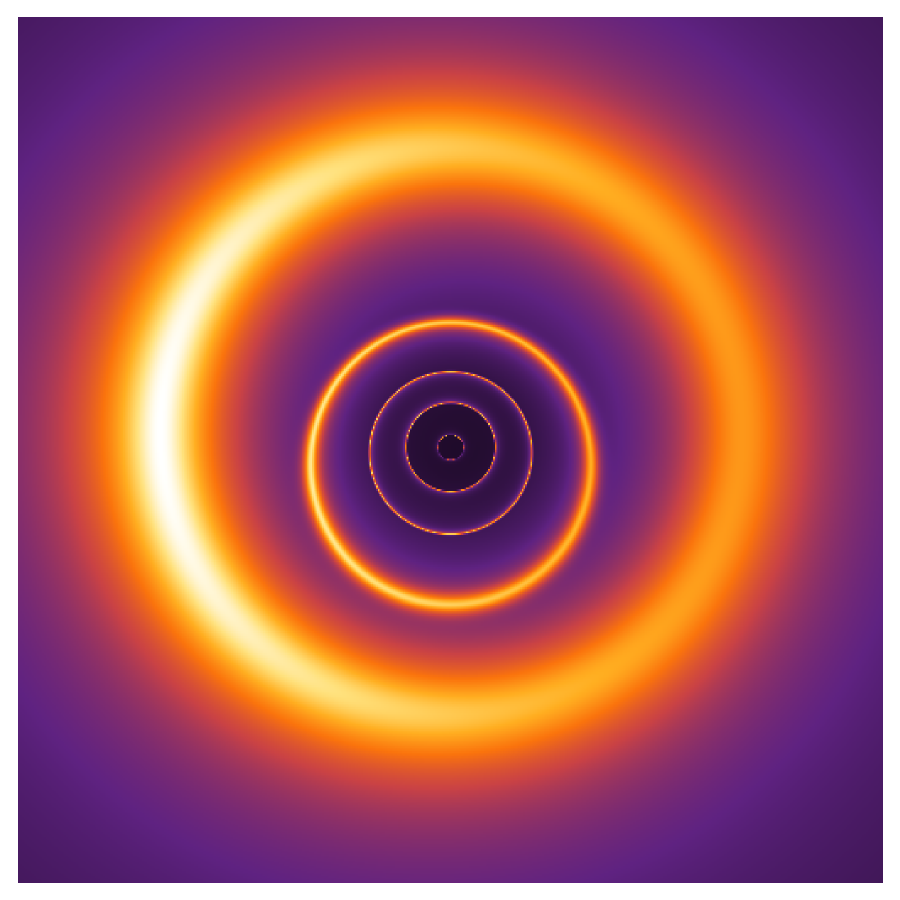}}
}
\subfigure[]{
\setcounter{subfigure}{0}
\subfigure[$\Lambda=200,\phi_0=0.06$]{\includegraphics[width=.24\textwidth]{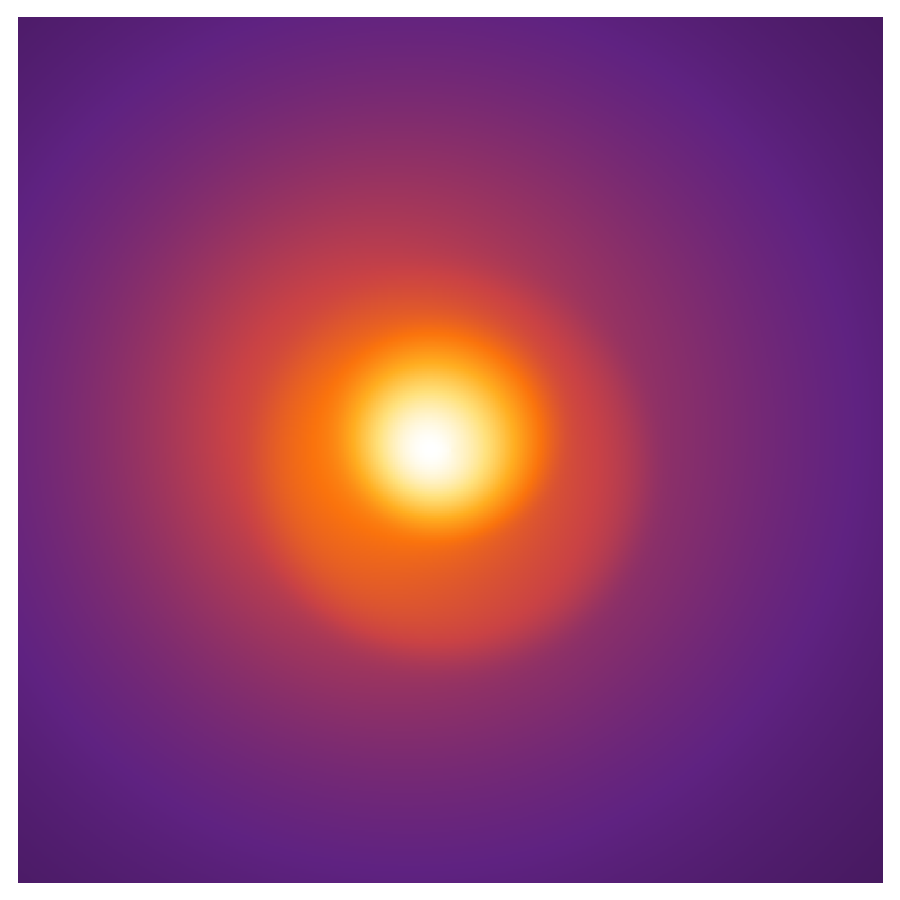}}
\subfigure[$\Lambda=200,\phi_0=0.09$]{\includegraphics[width=.24\textwidth]{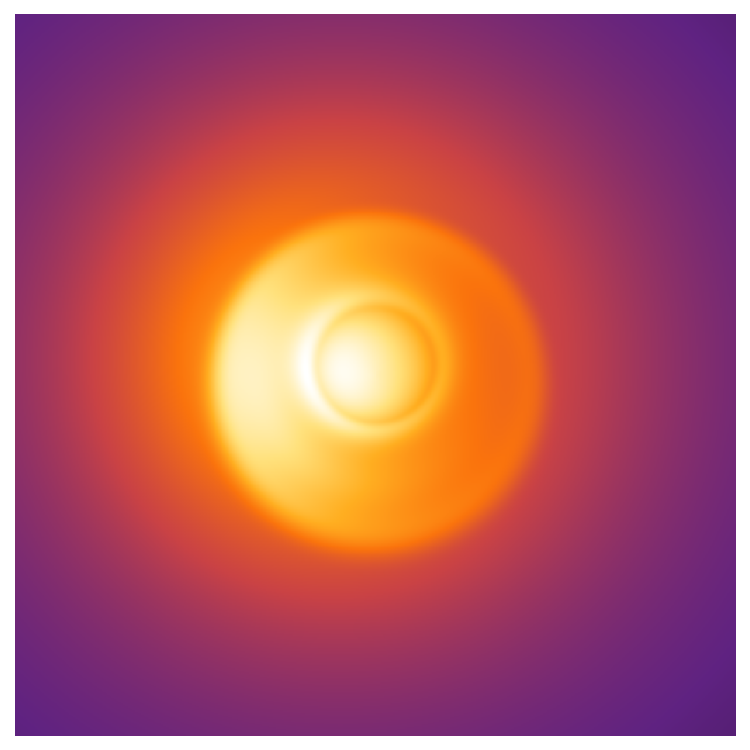}}
\subfigure[$\Lambda=200,\phi_0=0.15$]{\includegraphics[width=.24\textwidth]{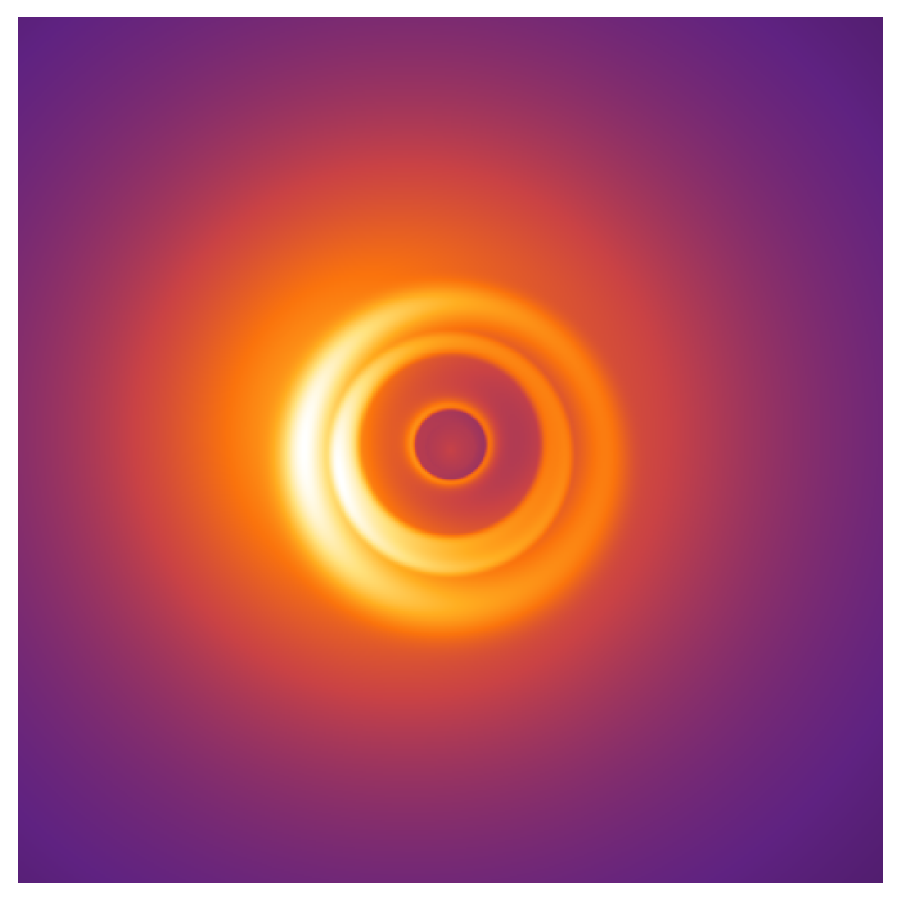}}
\subfigure[$\Lambda=200,\phi_0=0.18$]{\includegraphics[width=.24\textwidth]{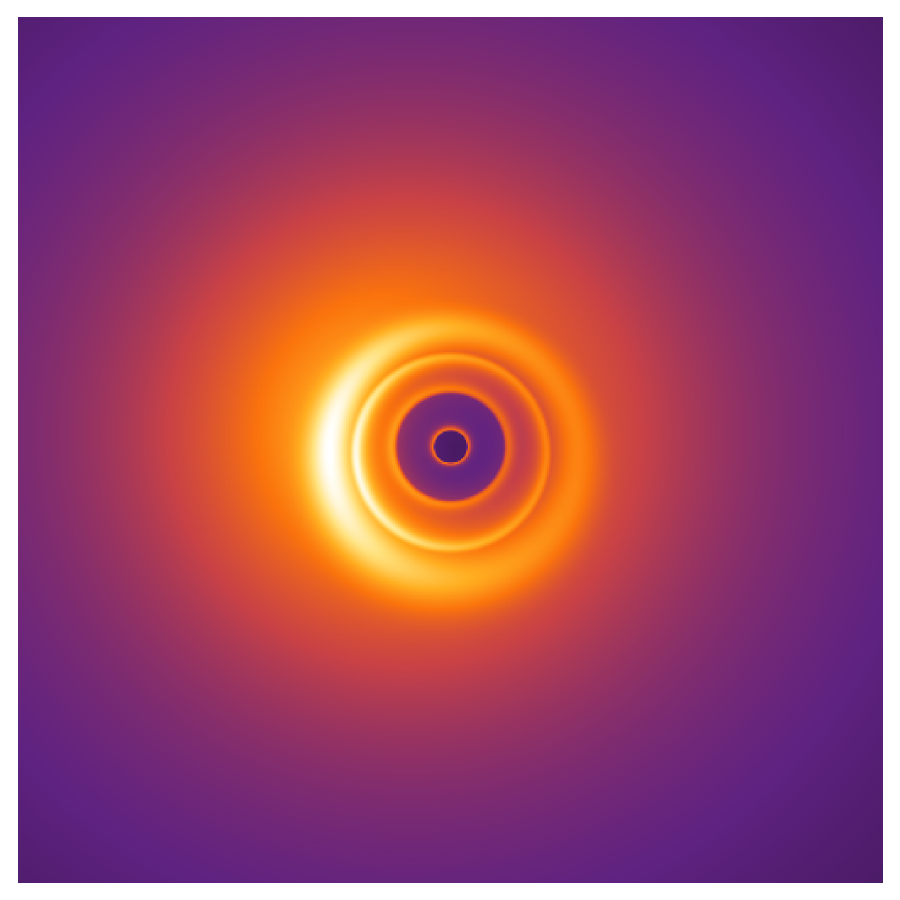}}
}
\vspace{-0.2cm}
\subfigure[]{
\setcounter{subfigure}{0}
\subfigure[$\Lambda=100,\phi_0=0.15$]{\includegraphics[width=.24\textwidth]{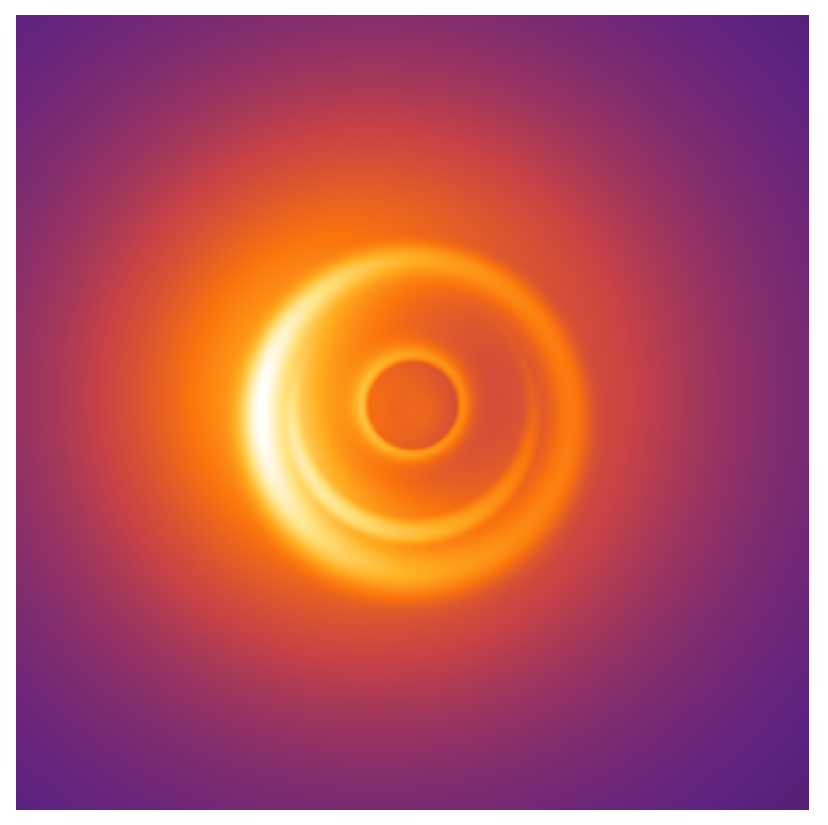}}
\subfigure[$\Lambda=200,\phi_0=0.15$]{\includegraphics[width=.24\textwidth]{200015m-17d.pdf}}
\subfigure[$\Lambda=300,\phi_0=0.15$]{\includegraphics[width=.24\textwidth]{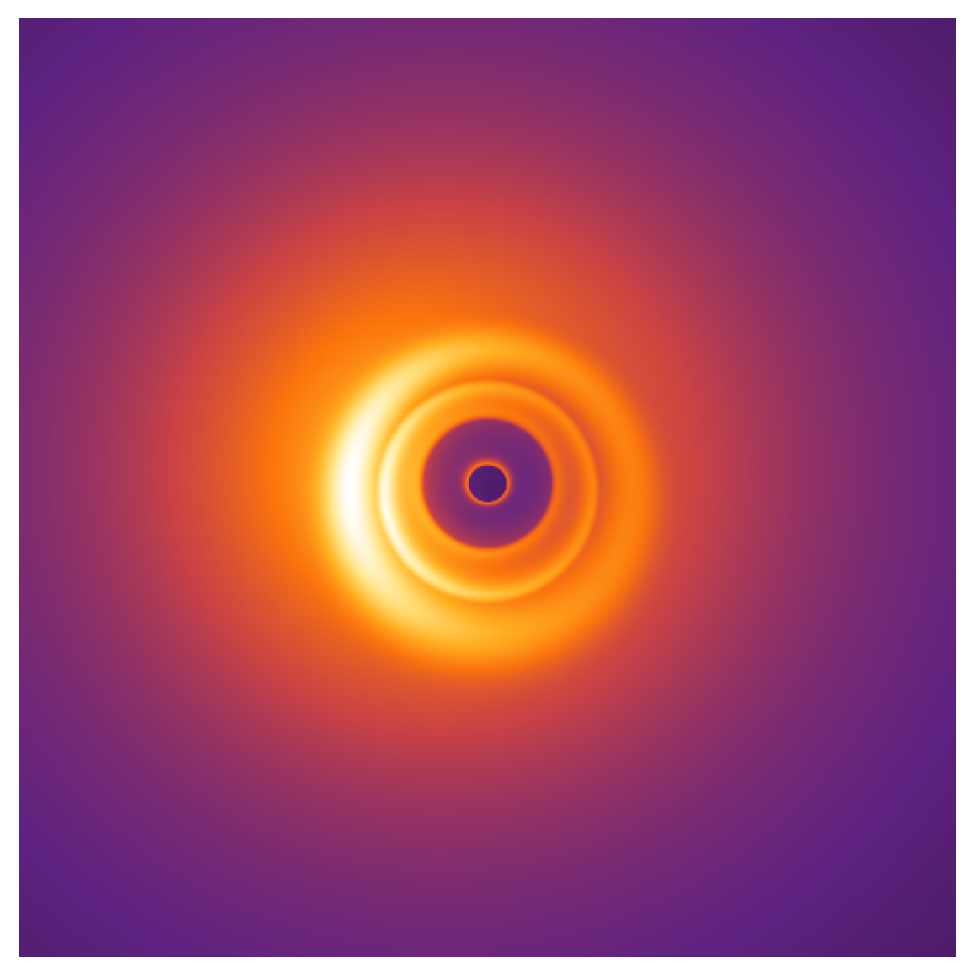}}
\subfigure[$\Lambda=500,\phi_0=0.15$]{\includegraphics[width=.24\textwidth]{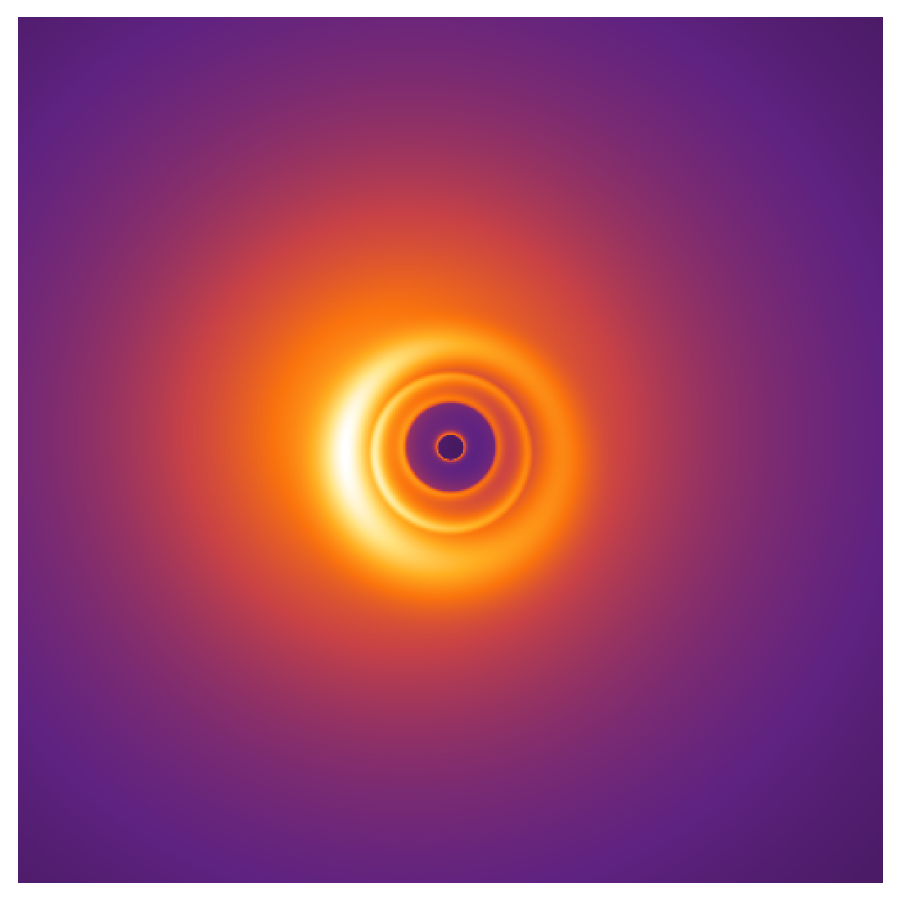}}
}
\vspace{-0.2cm}
\caption{\label{figbp} The images of boson stars, where $\theta_{obs}=17^\circ$, where rows 1-2: Model a; rows 3-4: Model b.}
\end{figure}

Through a simple calculation, we found that the ISCOs of the boson stars involved in this paper are all very small or even nonexistent. Therefore, for{ Model a}, we set the peak of this emission model at the position $6M$, with the intensity gradually decaying to zero on either side of $6M$. For{ Model b}, the peak of the emission model is set at $r=0$. The main purpose of selecting Model b is to compare the imaging results with those of black holes, because the accretion disk of a black hole extends at most to the event horizon ($2M$). For{ Models a} and {b}, we obtained images of boson stars, which are shown in Figures \ref{figbp} and \ref{figbpm}, respectively.

By using{ Model a}, the observable images of boson stars are presented in Figure \ref{figbp}\footnote{Notably, although we provided ten boson stars in the previous tables and figures, this figure mainly displays images of eight boson stars, as some of them appear quite similar. For details of those boson stars, please refer to Tables 1 and 2. } .
It first shows that the light rings in the boson star images-particularly the direct emission-exhibit slightly higher brightness on the left side compared to the right, which results from the redshift effect induced by the accretion disk with a four-velocity.
From the first image in the first row of Figure \ref{figbp}, we can see that this boson star image contains two parts: the larger ring is the direct image of the accretion disk, while another part is a bright crescent-shaped pattern, which is a lensing image. For the next three images in the first row, we observe that the number of light rings in those images increases as the parameter $\phi_0$ increases. For instance, when $\phi_0=0.09$, the image includes three light rings; when $\phi_0=0.15$ or $0.18$, the image shows five light rings, with the positions and spacings of those rings differing between the two images.
For the case of $\Lambda=200, \phi_0=0.09$, the outermost ring of this boson star being the direct image and the other two being lensing images. For the cases of $\Lambda=200, \phi_0= 0.15$ or $\phi_0=0.18$, the image contains five rings, where the outermost ring is the direct image. The second and fifth rings (from outside to inside) are lensing images, while the third and fourth rings are third-order images. Additionally, the results show that there is a black region in the center of the light ring, resembling the boson star shadow, and the size of this region gradually decreases as $\phi_0$ increases.
From the second row of Figure \ref{figbp}, it can be seen that when the observer is at $\theta_{obs}=17^\circ$, the number of light rings of boson stars increases from 4 to 5 as $\Lambda$ increases, and then the size of rings becomes progressively smaller, i.e., the innermost light ring. Meanwhile, as $\Lambda$ a increases, the darker central shadow region progressively shrinks.

For{ Model b}, we also obtained the images of boson stars, which are shown in Figure \ref{figbp}.
Based on the third row of Figure \ref{figbp}, it can be seen that since photons are emitted from the innermost edge of the accretion disk $r=0$, the image exhibits a rather distinctive ``Central
Emission Region" pattern, i.e., the first two figures in this row. And, the black-hole-like shadow of boson stars disappears, which indicates that they are differ from the shadow features of black holes(the lack of an event horizon). This phenomenon constitutes a significant observational distinction between these two classes of compact objects.
In addition, it can be seen that the direct image and the lensing image are almost distinguishable, as well as the ring-like structure is no longer observable.
However, as the parameters $\phi_0$ increase, the spacetime structure undergoes some changes, and the images of boson stars now exhibit two light rings as well as a darker central region.
And, the larger $\phi_0$, the more pronounced the central dark region becomes.
In this case, the direct emission and higher-order rings gradually become distinguishable for the last two panels in the third row.
For different values of $\Lambda$, similar to $\phi_0$, as $\Lambda$ increases, the dark region in the boson star image becomes more pronounced, and the higher-order subrings in the center grow increasingly distinct. Also, the size of light rings of the boson star decreases when $\Lambda$ increase.
As $\Lambda$ increase, boson stars progressively develop black-hole-like shadow features, a phenomenon that may lead to greater resemblance to black hole imaging characteristics in observational data.

For Schwarzschild black holes, the accretion disk is fundamentally constrained by the event horizon, with its innermost edge terminating at $2M$. In contrast, boson stars¡ªlacking this critical structure-allow their disks to extend seamlessly to the central region ($r=0$). This intrinsic distinction gives rise to unique observational signatures, such as the prominent ``Central Emission Region" (in Fig. \ref{figbp}), a defining feature absent in black hole systems and a key diagnostic marker for horizonless compact objects.
Notably, even when the accretion disk is positioned farther from the center (e.g., at $6M$), boson stars also exhibit intricate higher-order subrings-nested, symmetrically arranged emission structures that encode essential physical information. These features, entirely unattainable in black hole environments, serve as another clear discriminant between the two classes of objects.

\section{Polarized images of boson stars}\label{sec4}
\begin{figure}[htp]
\makeatletter
\renewcommand{\@thesubfigure}{\hskip\subfiglabelskip}
\makeatother
\centering
\subfigure[]{
\setcounter{subfigure}{0}
\subfigure[$\Lambda=200,\phi_0=0.06$]{\includegraphics[width=.24\textwidth]{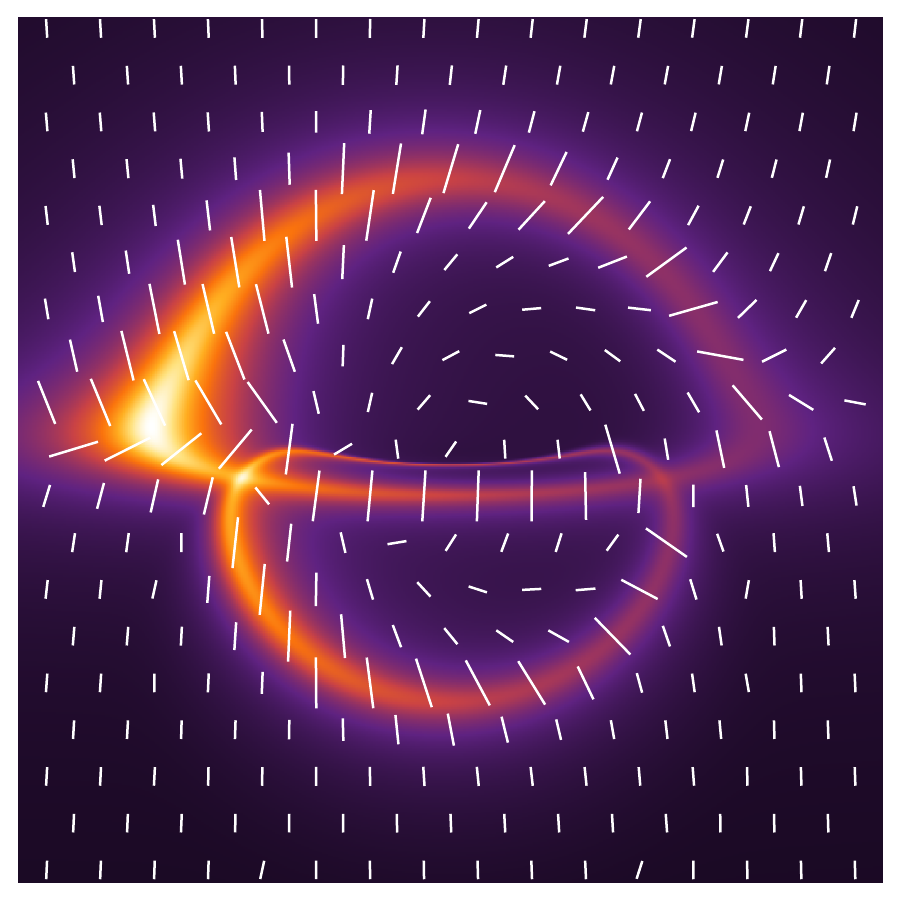}}
\subfigure[$\Lambda=200,\phi_0=0.09$]{\includegraphics[width=.24\textwidth]{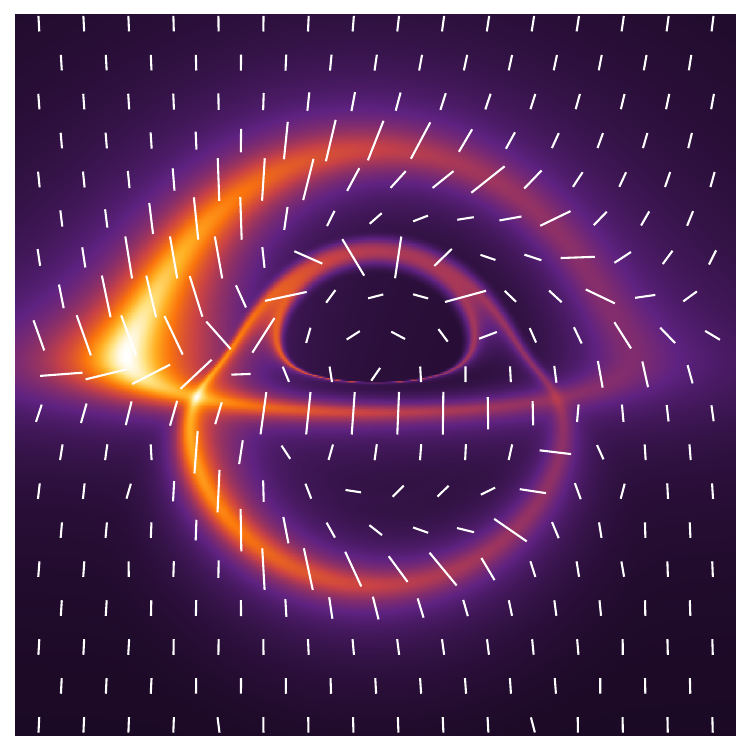}}
\subfigure[$\Lambda=200,\phi_0=0.15$]{\includegraphics[width=.24\textwidth]{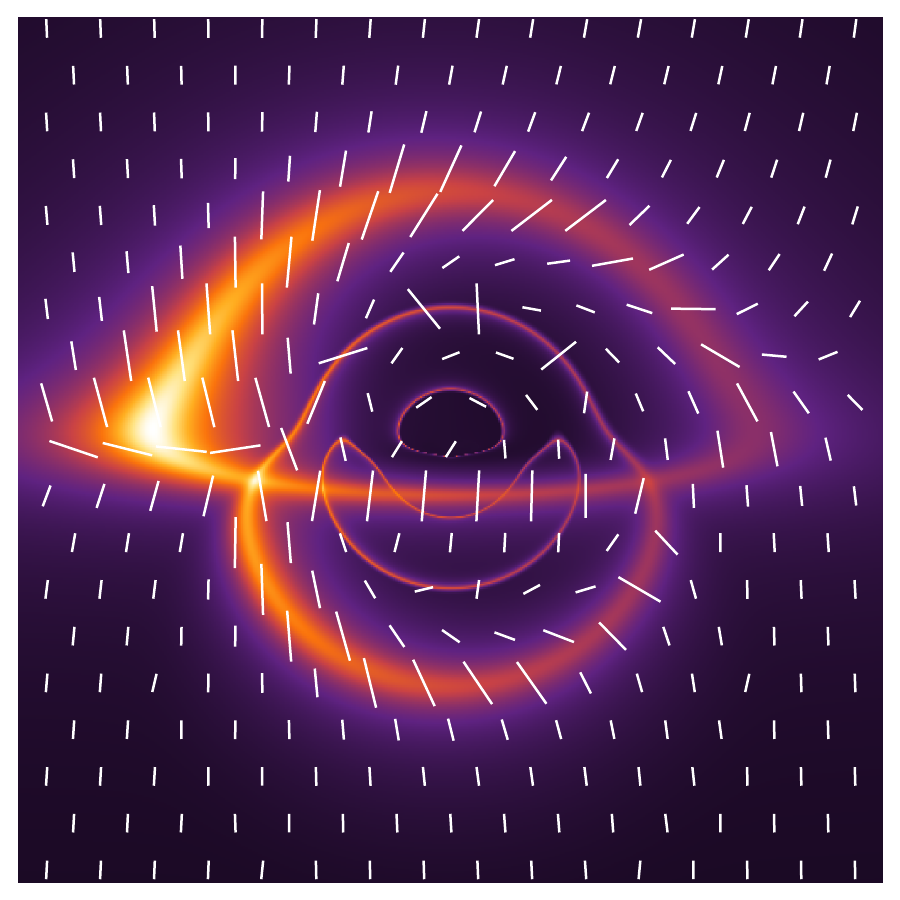}}
\subfigure[$\Lambda=200,\phi_0=0.18$]{\includegraphics[width=.24\textwidth]{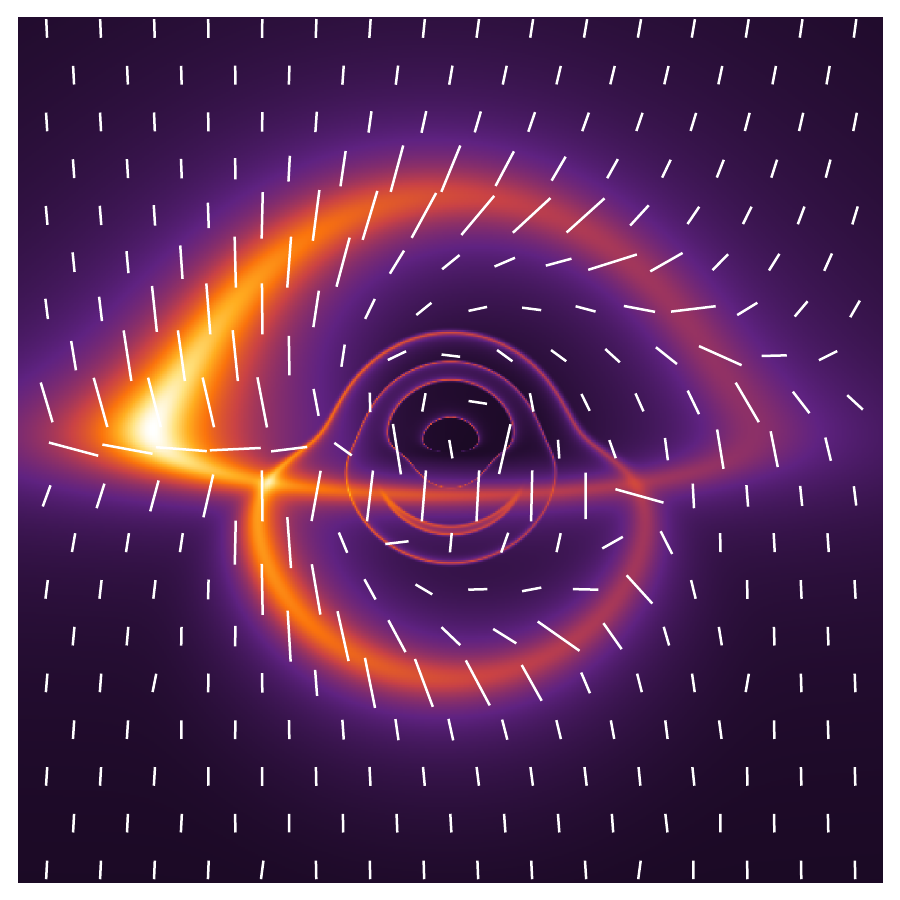}}
}
\subfigure[]{
\setcounter{subfigure}{0}
\subfigure[$\Lambda=100,\phi_0=0.015$]
{\includegraphics[width=.24\textwidth]{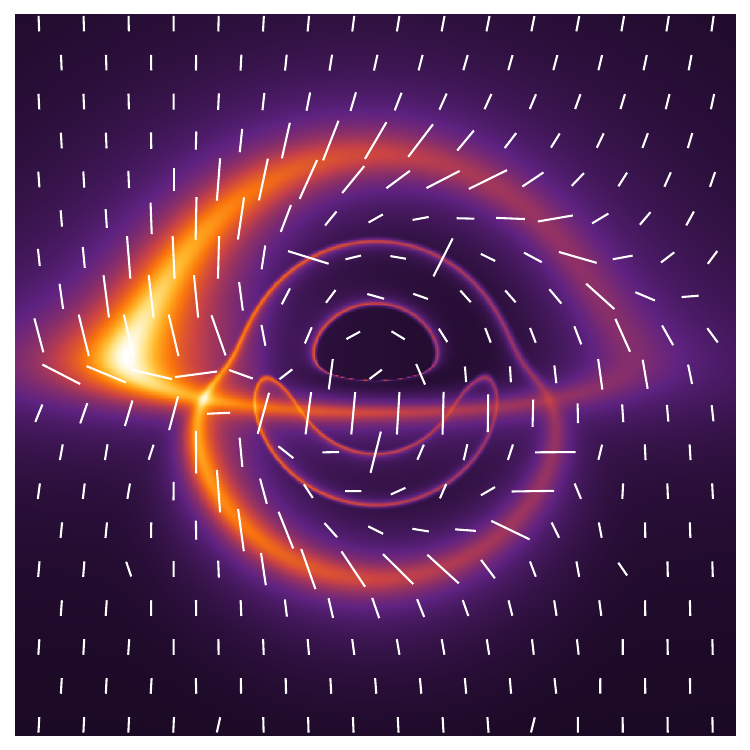}}
\subfigure[$\Lambda=200,\phi_0=0.15$]
{\includegraphics[width=.24\textwidth]{jihua200015-80d-5.pdf}}
\subfigure[$\Lambda=300,\phi_0=0.15$]{\includegraphics[width=.24\textwidth]{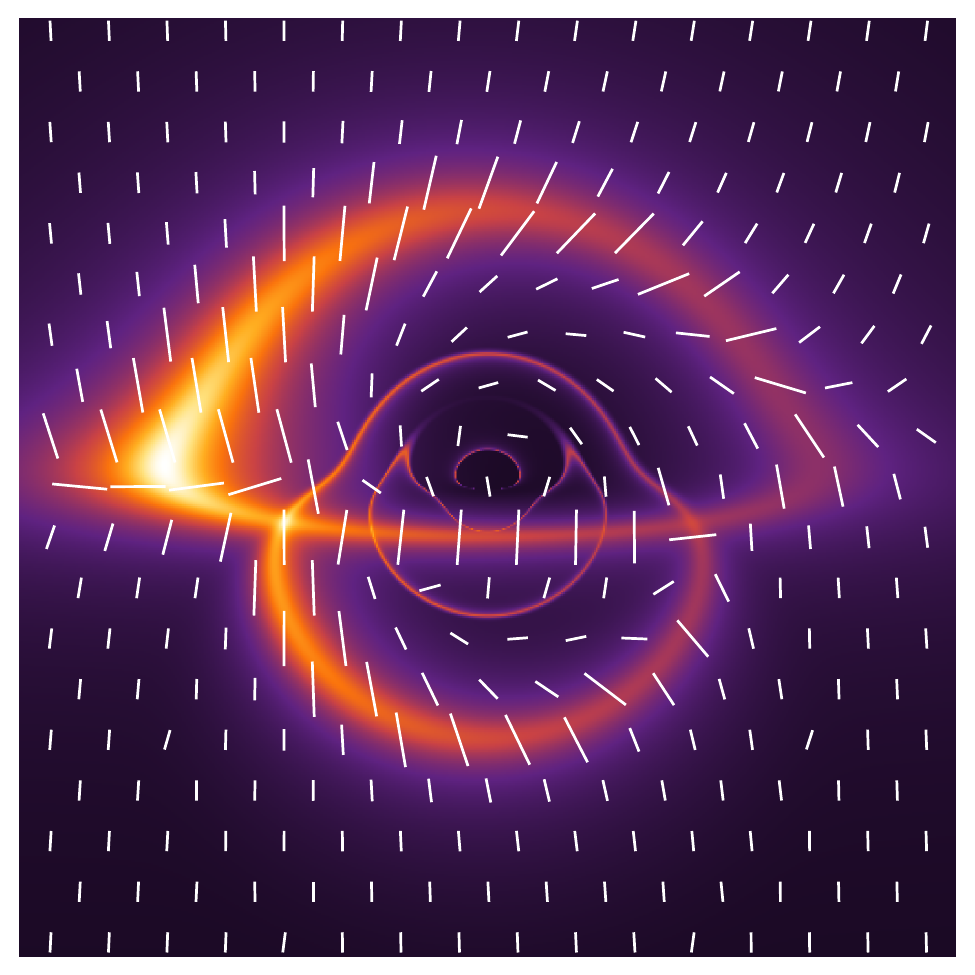}}
\subfigure[$\Lambda=500,\phi_0=0.15$]{\includegraphics[width=.24\textwidth]{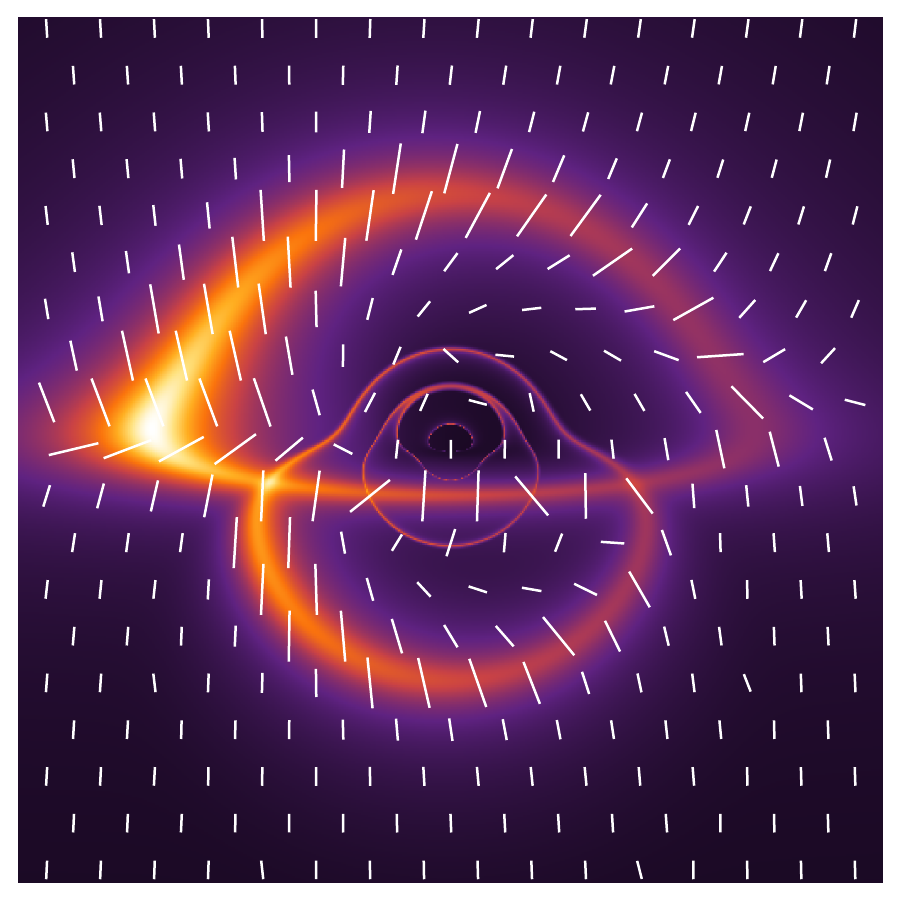}}
}
\subfigure[]{
\setcounter{subfigure}{0}
\subfigure[$\Lambda=200,\phi_0=0.06$]{\includegraphics[width=.24\textwidth]{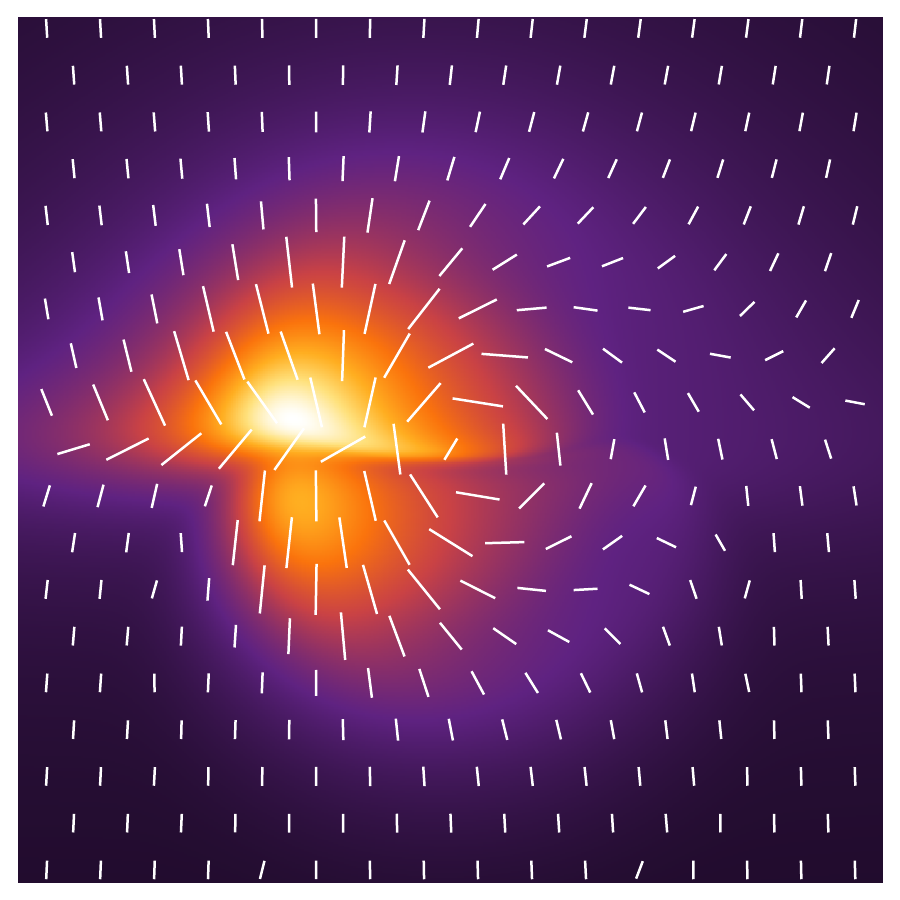}}
\subfigure[$\Lambda=200,\phi_0=0.09$]{\includegraphics[width=.24\textwidth]{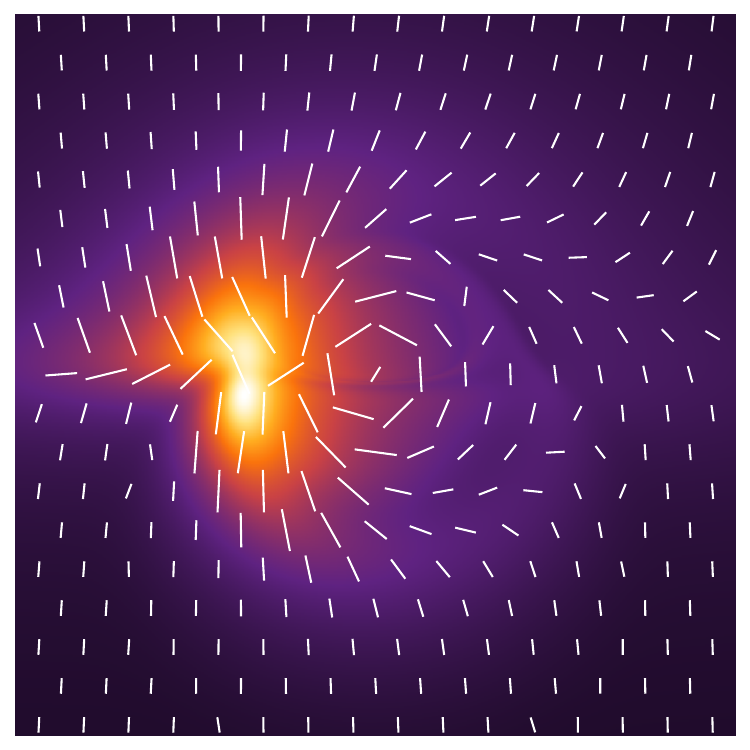}}
\subfigure[$\Lambda=200,\phi_0=0.15$]{\includegraphics[width=.24\textwidth]{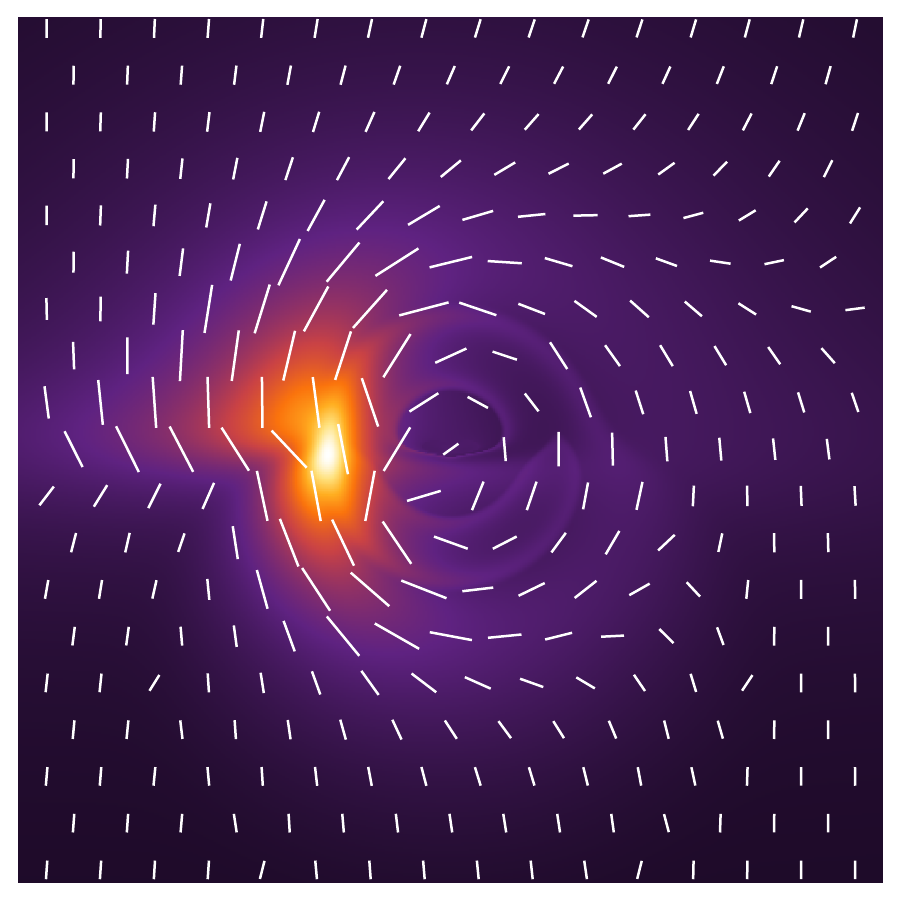}}
\subfigure[$\Lambda=200,\phi_0=0.18$]{\includegraphics[width=.24\textwidth]{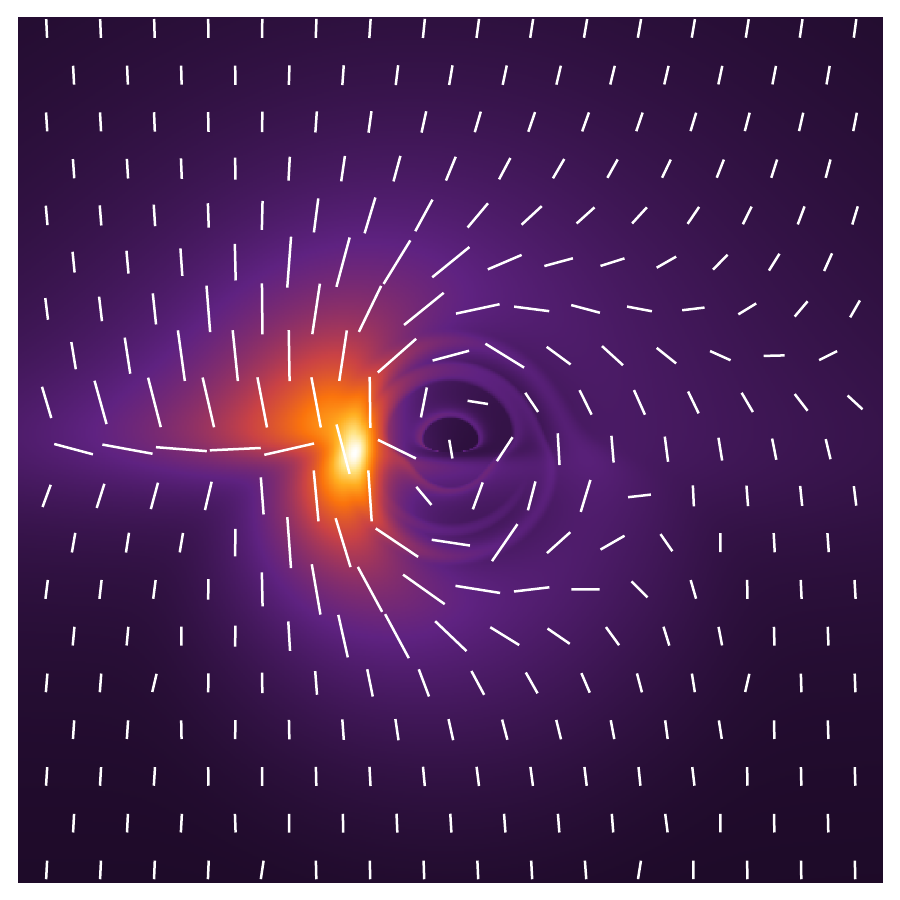}}
}
\subfigure[]{
\setcounter{subfigure}{0}
\subfigure[$\Lambda=100,\phi_0=0.15$]{\includegraphics[width=.24\textwidth]{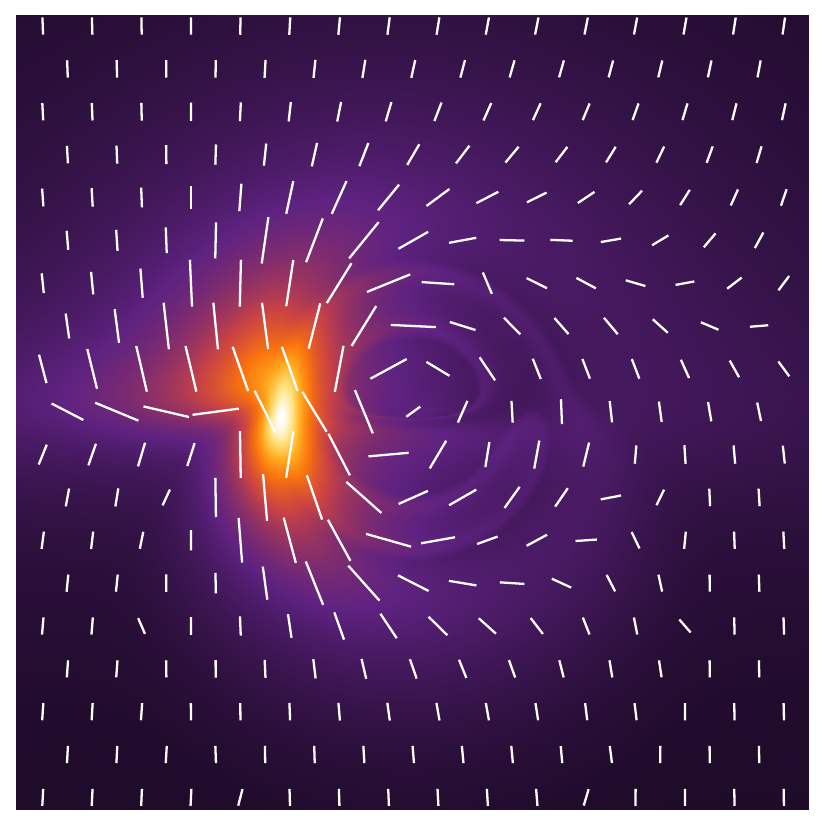}}
\subfigure[$\Lambda=200,\phi_0=0.15$]{\includegraphics[width=.24\textwidth]{jihua200015m-80d-5.pdf}}
\subfigure[$\Lambda=300,\phi_0=0.15$]{\includegraphics[width=.24\textwidth]{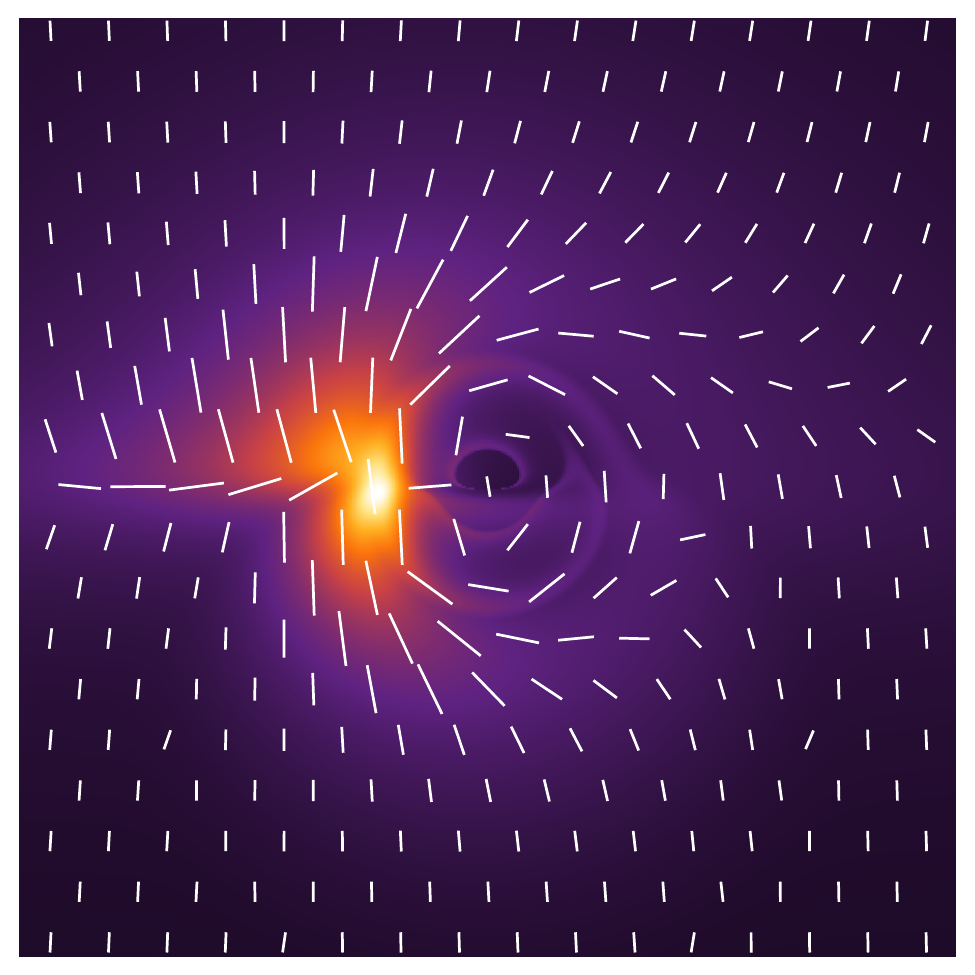}}
\subfigure[$\Lambda=500,\phi_0=0.15$]{\includegraphics[width=.24\textwidth]{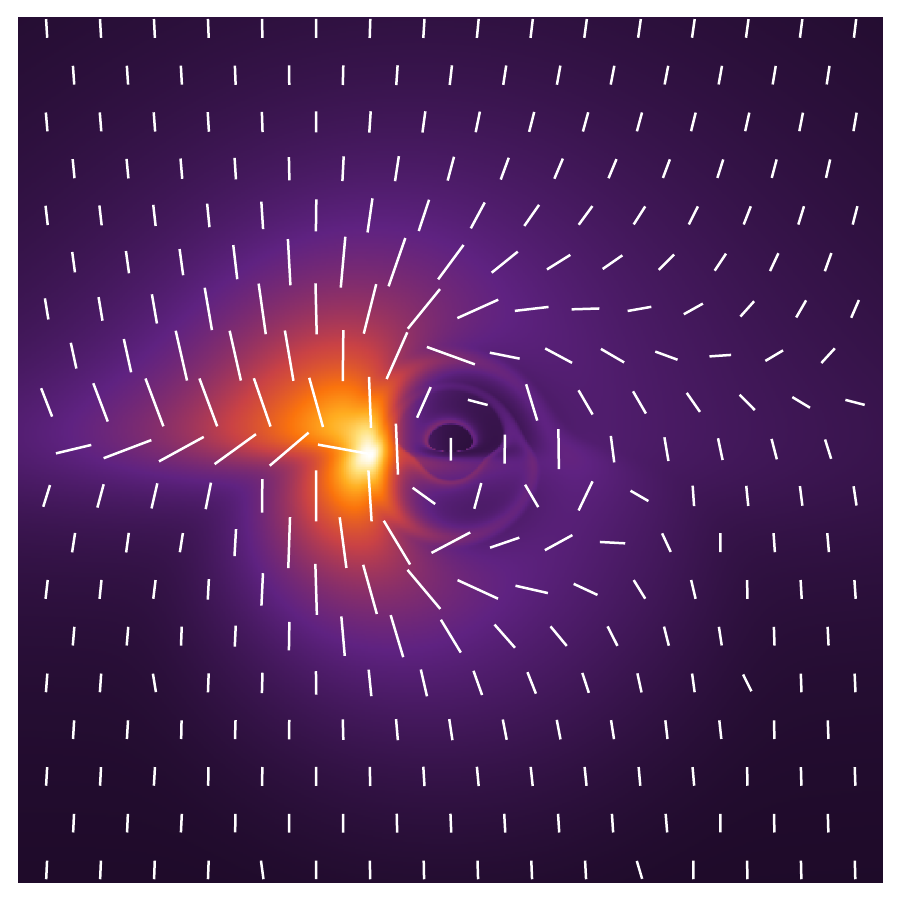}}
}
\vspace{-0.2cm}
\caption{\label{figpzbp} The polarized intensity tick plots in the total observed image of boson stars for $\theta_{obs}=80^{\circ}$, where rows 1-2: Model a; rows 3-4: Model b.}
\end{figure}

In the previous section, we thoroughly analyzed the accretion disk images of boson stars, revealing the variation of these images with respect to the initial scalar potential and coupling parameters. Based on these findings, this section will further investigate the characteristics of polarized images of boson stars, aiming to provide a more comprehensive understanding of their radiative properties and interactions with the surrounding environment.
Here, we assume that the emission of polarized light originates from synchrotron radiation emitted by electrons within a plasma.
As seen by an observer comoving with the plasma, $u^{\mu}$, the polarization direction $\vec{f}$ of the emitted light is perpendicular to both the local magnetic field $\vec{B}$ and the wave vector $\vec{k}$ of the photon, which is
\begin{equation}
\vec{f} = \frac{\vec{k} \times \vec{B}}{|\vec{k}||\vec{B}|}.
\label{eq:vector}
\end{equation}
The generally covariant form of Eq.(\ref{eq:vector}) can be expressed as,
\begin{equation}
f^{\mu} \propto \epsilon^{\mu\nu\rho\sigma} u_{\nu} k_{\rho} B_{\sigma}.
\label{eq:vector1}
\end{equation}
with $u^{\mu}$ is the four-velocity of the fluid, $k^{\mu}$ represents the four-wavevector of the photon and $B^{\mu}$ is the magnetic field. In general, we can first find the direction of the polarization vector and then normalize it so that it satisfies orthonormality, ie., $f^{\mu}f_{\mu}=1$.
The intensities of linearly polarized light and natural light at the emission site can be represented by the emission functions $J_p$ and $J_i$\footnote{Here, $J_i$ corresponds to the emissivity (\ref{jn}).}, respectively. For simplicity, we assume that the emission intensity is independent of the photon frequency and magnetic field, and depends solely on the position. So, we have
\begin{equation}
J_i=J_i{(r)}, \quad J_p=\varpi J_i(r),
\label{eq:vector3}
\end{equation}
where $\varpi$ is a function used to describe the proportion of linearly polarized light in the total light intensity at the emission point, $\varpi \in [0, 1]$. And, if we assume that the emitted light is completely linearly polarized, then we can take $\varpi=1$.
Based on the geometric optics approximation, the polarization vector $f^{\mu}$ undergoes parallel transport along the geodesic of the photon, i.e.,
\begin{equation}
    k^\nu \nabla_\nu f^\mu = 0.
    \label{eq:placeholder_label}
\end{equation}
By considering the affine parameter $\lambda$, Eq.(\ref{eq:placeholder_label}) can also be rewritten as
\begin{equation}
    \frac{d}{d\lambda} f^\mu + \Gamma^\mu_{\nu\rho} k^\nu f^\rho = 0.
    \label{eq:polarization_evolution}
\end{equation}
At the observer's location, the intensity of the linearly polarized light $P_{v_o}$ and the total intensity $I_{v_o}$ are the same as in the unpolarized case, and can be expressed by the following equation,
\begin{equation}
    P_{v_o}= g^3 J_p, \quad I_{v_o}= g^3 J_i,
    \label{eq:intensity}
\end{equation}
with $g$ is redshift factor.
In subsection \ref{sec4}, we have established a set of orthogonal tetrads (ZAMO) at the observer's location, along with a screen for imaging the boson star. Based on the two orthogonal basis vectors ($e_{(\theta)}$, $e_{(\varphi)}$) in the screen plane, the projection of the polarization vector onto the coordinate plane can be expressed as,
\begin{equation}
    f^{(\alpha)} = f^{\mu} \cdot {e}_{\alpha}= - f^{\mu} \cdot {e}_{\varphi}, \quad f^{(\beta)} = f^{\mu} \cdot {e}_{\beta} = - f^{\mu} \cdot {e}_{\theta}.
\end{equation}
Here, we have chosen the gauge $f^{(\beta)} >0$, $\chi \in (0, \pi)$ during the numerical calculation.
After choosing the direction of the polarization vector, we can obtain the total intensity of the linearly polarized light observed by the observer by summing the linearly polarized light emitted at each emission point on the equatorial plane, which gives the total intensity of the linearly polarized light. According to the definitions of the Stokes parameters $Q$ and $U$\cite{Huang:2024bar}, these parameters satisfy the principle of linear superposition. Thus, the final result can be obtained by direct summation of $Q$ and $U$, which gives
\begin{equation}
Q_{all}=\sum_{n=1}^{n_{max}} g_n^3 J_{p_n}
\left((f_n^{(\alpha)})^2-(f_n^{(\beta)})^2\right), \quad
U_{all}=\sum_{n=1}^{n_{max}} g_n^3 J_{p_n}\left(2(f_n^{(\alpha)}f_n^{(\beta)}\right).
\label{Etoal}
\end{equation}
In Eq.(\ref{Etoal}), $n$ denotes the number of times the light intersects the equatorial plane, $n_{max}$ represents the maximum number of intersections. Thus, the total intensity of the linearly polarized light is
\begin{equation}
    P_o = \sqrt{Q_{all}^2+U_{all}^2}.
\end{equation}
The electric vector position angle (EVPA) reads
\begin{equation}
    \chi = \dfrac{1}{2}\arctan{\dfrac{U_{all}}{Q_{all}}}.
\end{equation}
Therefore, when we adopt the accretion disk model from Section \ref{sec4}, we can obtain the direction and intensity of the polarization vector at the observer's location. By considering the thin disk used in previous sections, the polarized image of boson stars is presented in Figure \ref{figpzbp}, where the observer's position locates at $\theta_{obs} = 80^\circ$.

In light of the findings reported by EHT for $M87$*, where the optimal magnetic field configuration $B(r, \theta, \varphi) = (0.87, 0, 0.5)$ has been demonstrated to best account for the observed polarization features, this study will employ identical parameters to simulate the polarized emission for boson stars in Figure \ref{figpzbp}.
For $\Lambda=200, \phi_0=0.15$ in Model a, the image consists of four bright rings: (i), Outer arc-shaped ring: this is the outermost ring in the image, appearing as an arc-shaped structure covering the upper part of the image.
(ii), Second outer symmetrical ring: located below the first arc-shaped ring, this ring appears as a closed, symmetrical circle symmetric about the vertical direction.
(iii), Inner curved ring: This ring is in the middle-lower part of the image, taking on a symmetrical ``horseshoe" or ``crescent" shape.
(iv), Innermost small ring: in the center of the image, there is a smaller ring that is almost completely closed, resembling a small semicircle.
This structural characteristic is replicated similarly in Model b. However, due to the accretion disk's radiation emanating from $r=0$, these features become partially obscured by direct emission, making them difficult to observe. Consequently, the boson star images under Model b exhibit a left-centered radiant patch that demonstrates the progressive diminution with increasing parameter $\phi_0$, while maintaining the size invariance under parameter $\Lambda$ enhancement, during which the internal annular structures gradually emerge.

For the observed polarization features, it is obvious that the polarized intensity observed in the brighter regions of the accretion disk image (e.g., arc-shaped emission belts or photon rings) exhibits a marked enhancement compared to their darker counterparts (such as the region of shadow). For $\Lambda=200, \phi_0=0.15$ in Model a, the redshift effects induce a enhanced brightness in the left side of the direct emission ring compared to its right side, consequently generating a pronounced left-dominant polarization intensity distribution. Also, the polarized intensity of Second outer symmetrical ring significantly exceeds that of other higher-order subrings in the central region.
We also find that the polarized images emerge within the boson star's interior for both Model a and Model b. And, Model b exhibits a stronger polarization intensity than Model a, which is attributable to their different positions of accretion disk.
In black hole systems, the absence of escaping radiation within the event horizon theoretically precludes any observable polarization effects in their interior domains. However, we find the pronounced polarized emission features in boson stars' central regions, which provides a critical tool for distinguishing these two classes of compact objects.
In general, the intensity and direction of the total observed polarization are closely related to spacetime geometry, magnetic field, and accretion disk distribution. Therefore, we will further investigated the polarized images of boson stars under different spacetime parameters ($\phi_0$ and $\Lambda$), accretion disk's positions, and magnetic field configuration, with the numerical results presented in the following Figure \ref{figpz}.

\vspace{-0.1cm}
\begin{figure}[!h]
\makeatletter
\renewcommand{\@thesubfigure}{\hskip\subfiglabelskip}
\makeatother
\centering 
\vspace{-0.2cm}
\subfigure[$ $]{
\setcounter{subfigure}{0}
\subfigure[$(a)$]{\includegraphics[width=.23\textwidth]{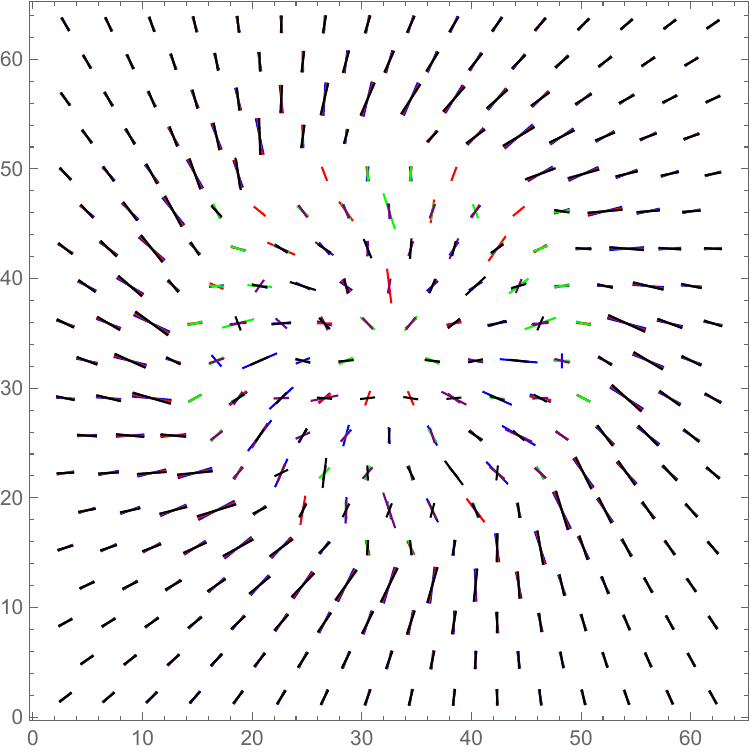}}
\subfigure[$(b)$]{\includegraphics[width=.23\textwidth]{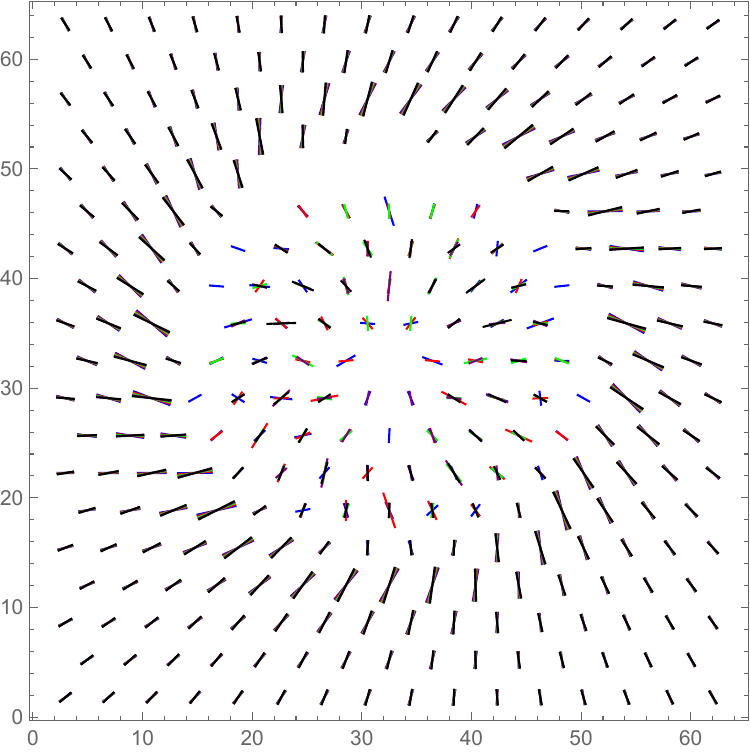}}
\subfigure[$(c)$]{\includegraphics[width=.23\textwidth]{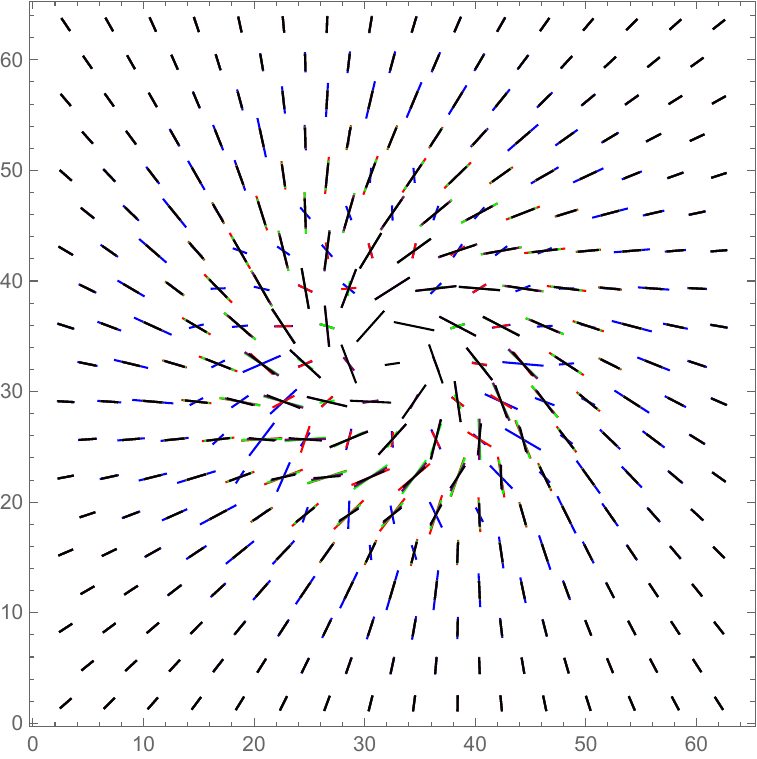}}
}
\subfigure[$(d)$]{\includegraphics[width=.23\textwidth]{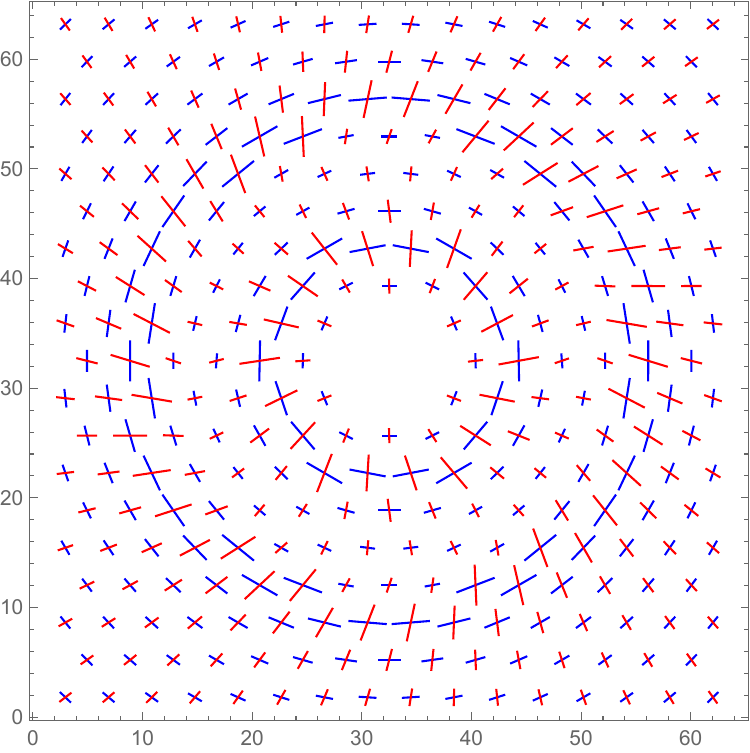}}
\vspace{-0.4cm}
\caption{\label{figpz} The total observed polarization versus boson stars' parameters, accretion disk's position and magnetic field. }
\end{figure}

In Figure \ref{figpz}, the blue, red, green, purple, and black line segments correspond to $\phi_0$=$0.06$, $0.09$, $0.12$, $0.15$, $0.18$ and $\lambda=200$ in subfigure (a). The results are based on the Model b in which $\beta=6M$ with $J_n=0$ when $r<6M$. The observer is positioned at $\theta=17^\circ$ and the direction of magnetic field is fixed as $B(r,\theta, \varphi) = (0.87, 0, 0.5)$, and the accretion disk is static.
Subfigure (b) examines the effect of varying $\Lambda$, with the line segments representing $\Lambda=100, 200, 300, 400, 500$(blue, red, green, purple, and black, respectively.), while keeping $\phi_0=0.15$. All other conditions remain identical to those in subfigure (a).
In subfigure (c), we fix $\phi_0=0.06$ and $\Lambda=200$ and explore the influence of disk's position. The line segments correspond to the inner edges of the disk at ($6M, 3M, 2M, M, 0$) (blue, red, green, purple, and black, respectively).
Additionally, the red and blue line segments in subfigure(d) are related to two magnetic field configurations: $B=(1, 0, 0)$ and $B=(0.87,0,0.5)$, respectively, with the observer now at $\theta=0^\circ$ for $\phi_0=0.06$ and $\Lambda=200$. All other parameters remain unchanged.
From the subfigures (a) and (b), we can see that the polarization intensity and direction only show some minimal variations with different values of the coupling parameter $\Lambda$ and initial scalar field $\phi_0$ in regions distant from the boson star's center. However, near the core of the boson star, these polarization characteristics (both intensity and direction) exhibit a very strong dependence on these two parameters.
As shown in Figure \ref{figpz}, the longest segments (for all colors) consistently appear near the region of maximum intensity of the direct emission ring, at which the differences in polarization magnitude and direction among different boson stars are not yet clearly evident.
But near the center of boson star, we can observe that the red line segments are significantly longer than those of other colors, while at other locations, blue or other colored segments dominate in length, which requires a magnification of the corresponding image for clear observation.
Different values of parameters ($\phi_0$ and $\Lambda$) correspond to different models of boson stars, consequently leading to some significant differences in their associated physical properties, including mass and radius.
And, when one is close to the boson stars, the gravitational field intensity is very strong, which make the differences in spacetime structure among boson stars more obvious.
As a result, the differences in morphology and position of lensing images and higher-order images formed by accretion disks around different boson stars will be more pronounced.
Not only that, since the polarization vector undergoes parallel transport along geodesics, these differences of space-time structure are still directly manifested in the intensity distribution and direction characteristics of the observed polarization image.
For subfigure(c), in the region far from the center of boson star, it is clear that,  the direction of polarization does not change with the position of the accretion disk, but its intensity changes with the shift of the accretion disk's position. However, within the interior region of boson star, as the accretion disk moves inward from $r=6M$ to $r=0$, both the polarization intensity and direction undergo obvious changes. This is because the central region is gradually covered by direct radiation, and the magnitude and direction of polarization require additional calculation of the flux of direct radiation.
The subfigure(d) shows that as the direction of the magnetic field changes, the polarization direction also exhibits a significant corresponding variation. On the observer's screen, the variation in polarization direction remains uniform at every position while the polarization magnitude stays unchanged.
Combined with above facts, it is true that the polarization images of boson stars can indeed reveal their spacetime geometry, the spatial distribution of accretion disks, and magnetic field configurations.

\section{In comparison with Schwarzschild black hole}\label{sec5}
\begin{figure}[!h]
\makeatletter
\renewcommand{\@thesubfigure}{\hskip\subfiglabelskip}
\makeatother
\centering 
\vspace{-0.2cm}
\subfigure[$ $]{
\setcounter{subfigure}{0}
\subfigure[$(a)$ static flow]{\includegraphics[width=.27\textwidth]{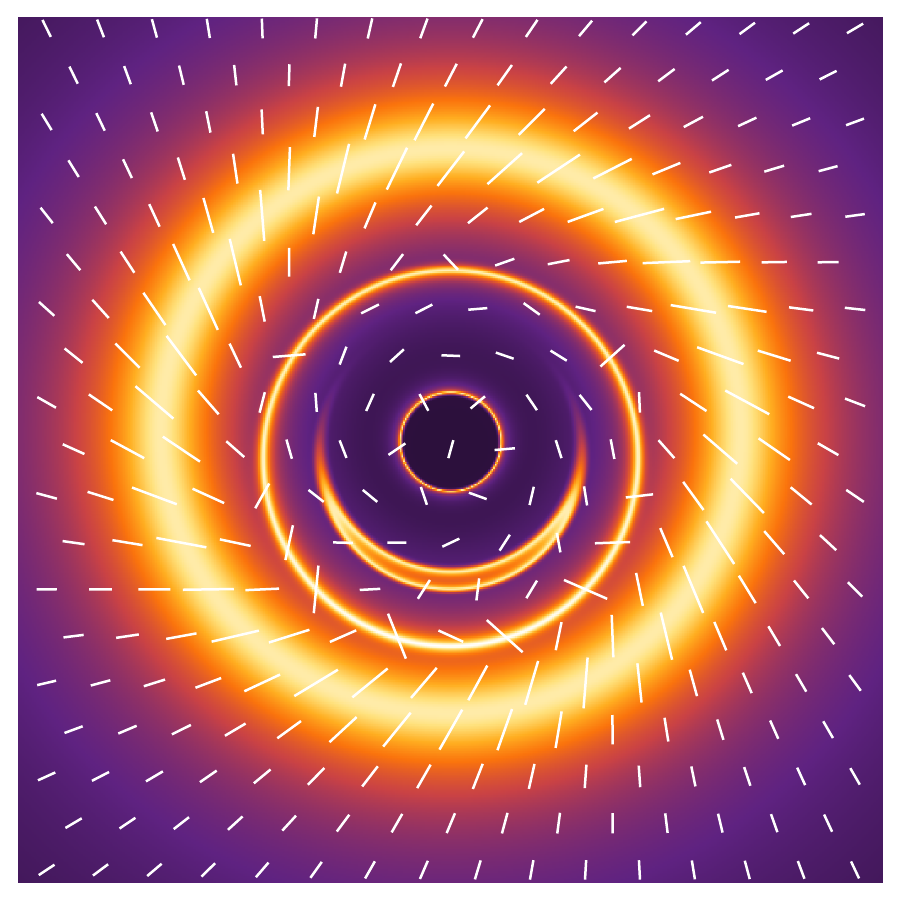}}
\subfigure[$(b)$ anticlockwise flow]{\includegraphics[width=.27\textwidth]{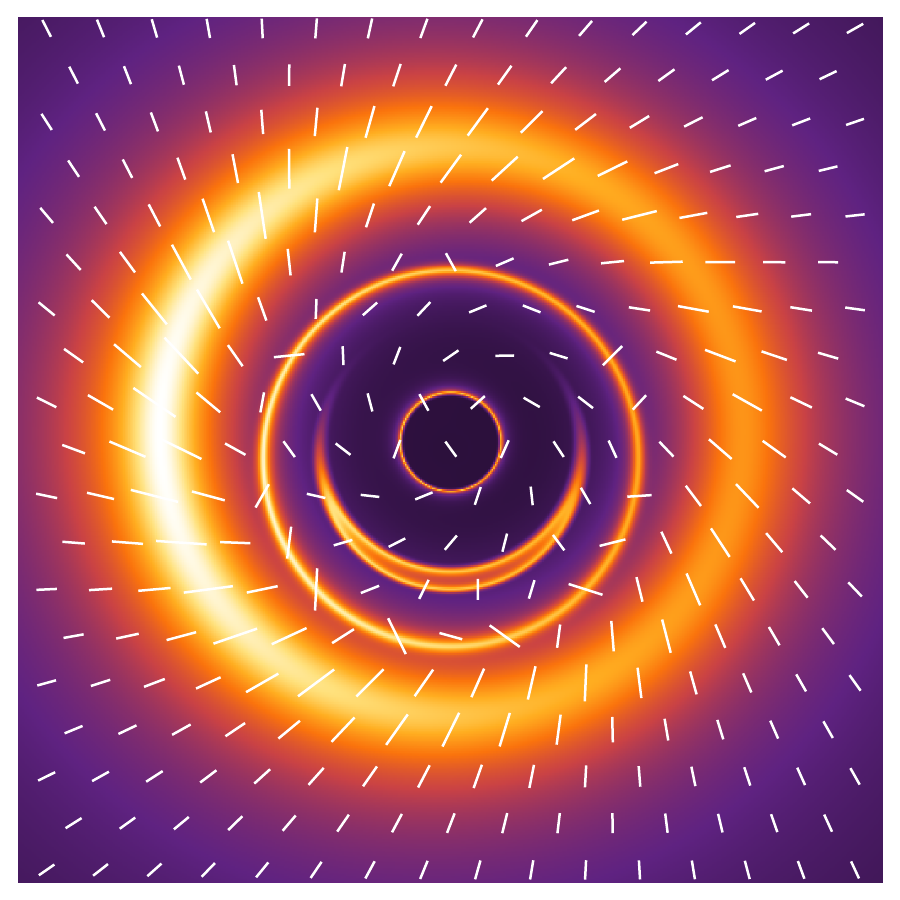}}
\subfigure[$(c)$ clockwise flow]{\includegraphics[width=.27\textwidth]{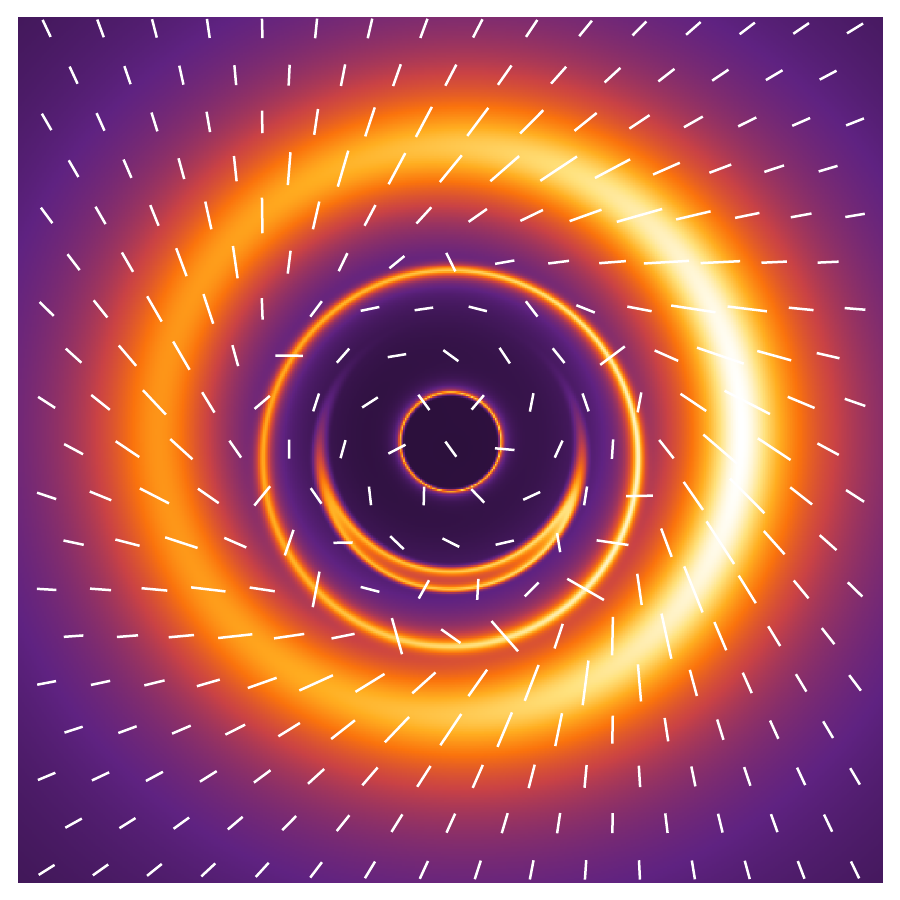}}
}
\vspace{-0.4cm}
\caption{\label{fig11} The images of boson stars, where the accretion disk is static, anticlockwise and clockwise flows.}
\end{figure}

In this section, before comparing with that of Schwarzschild black hole, we first examine how the accretion flow orientation affects the image of boson star.
Considering that the accretion disk is either static or rotating clockwise\footnote{The direction is defined as the accretion flow observed when viewing the equatorial accretion disk from a top-down perspective along the boson star's north pole.}, the corresponding images for{ Model a} are presented in Figure \ref{fig11}, where $\Lambda=200, \phi_0=0.15$, $B=(0.87, 0, 0.5)$, and the observed angle  is fixed as $\theta_{o}=17^\circ$.
\vspace{-0.1cm}

When the accretion disk is static, the brightness along the light rings around boson star is well-distributed. However, when the accretion flow moves in a circular orbit either clockwise or counterclockwise, the distribution of the intensity becomes asymmetric. This indicates that if the position of accretion disk remains unchanged, the change of its state do not affect the position of boson star's light rings but only influence their brightness. Therefore, one can see that the position and size of boson star's light rings (the higher-order rings) are determined solely by the properties of boson star's spacetime and are independent of the state of accretion flow.
Furthermore, the polarization orientation remains independent of the accretion flow direction, whereas the polarized intensity is proportional to the brightness of the disk's image.

\begin{figure}[!h]
\makeatletter
\renewcommand{\@thesubfigure}{\hskip\subfiglabelskip}
\makeatother
\centering 
\vspace{-0.2cm}
\subfigure[$ $]{
\setcounter{subfigure}{0}
\subfigure[$(a)$ boson star with $\theta_{obs}=0^\circ$ ]{\includegraphics[width=.27\textwidth]{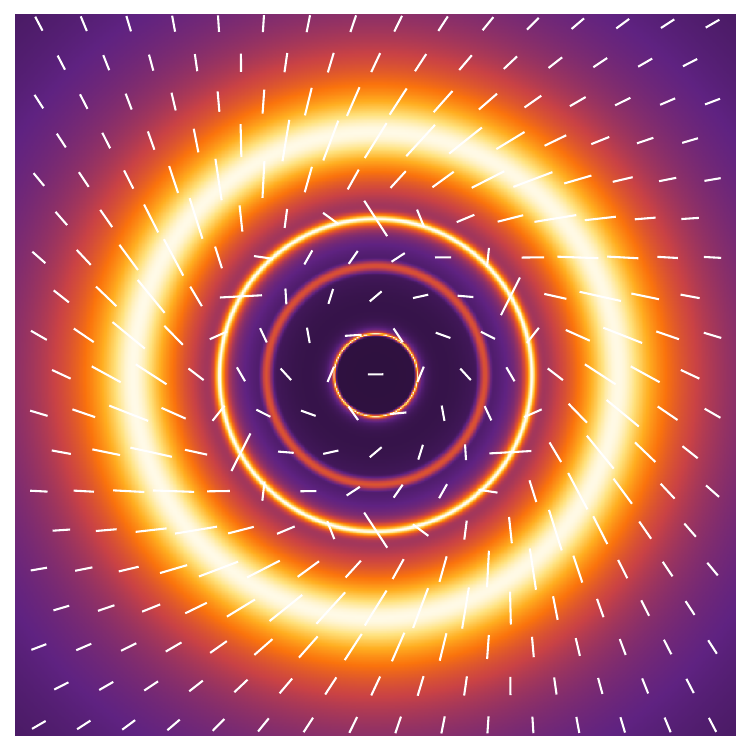}}
\subfigure[$(b)$ boson star with $\theta_{obs}=40^\circ$]{\includegraphics[width=.27\textwidth]{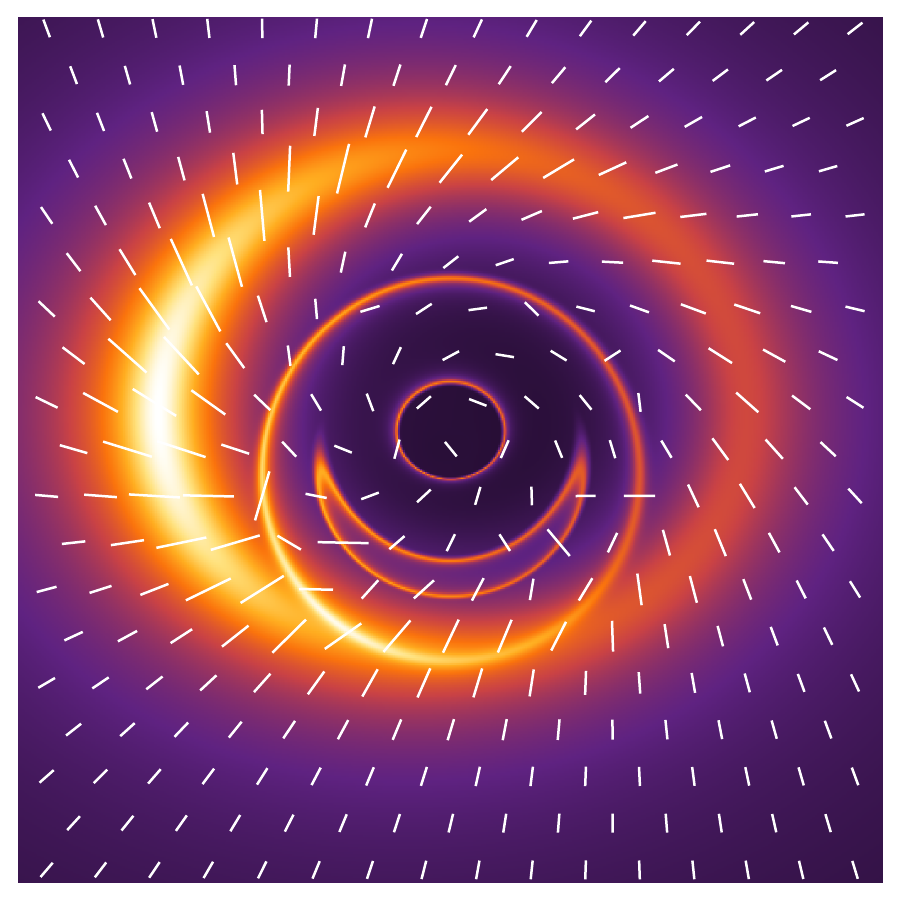}}
\subfigure[$(c)$ boson star with $\theta_{obs}=80^\circ$]{\includegraphics[width=.27\textwidth]{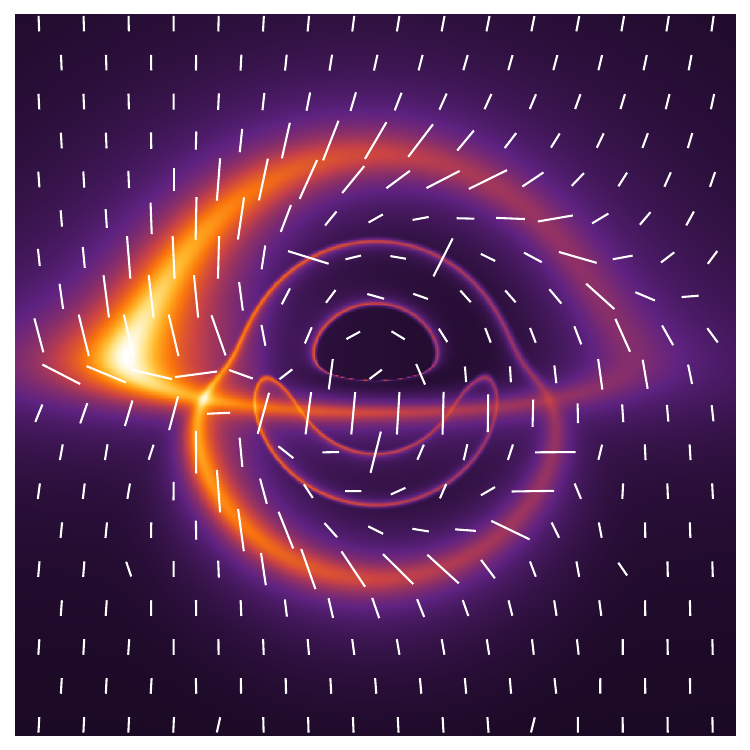}}
}
\subfigure[$ $]{
\setcounter{subfigure}{0}
\subfigure[$(d)$ black hole with $\theta_{obs}=0^\circ$ ]{\includegraphics[width=.27\textwidth]{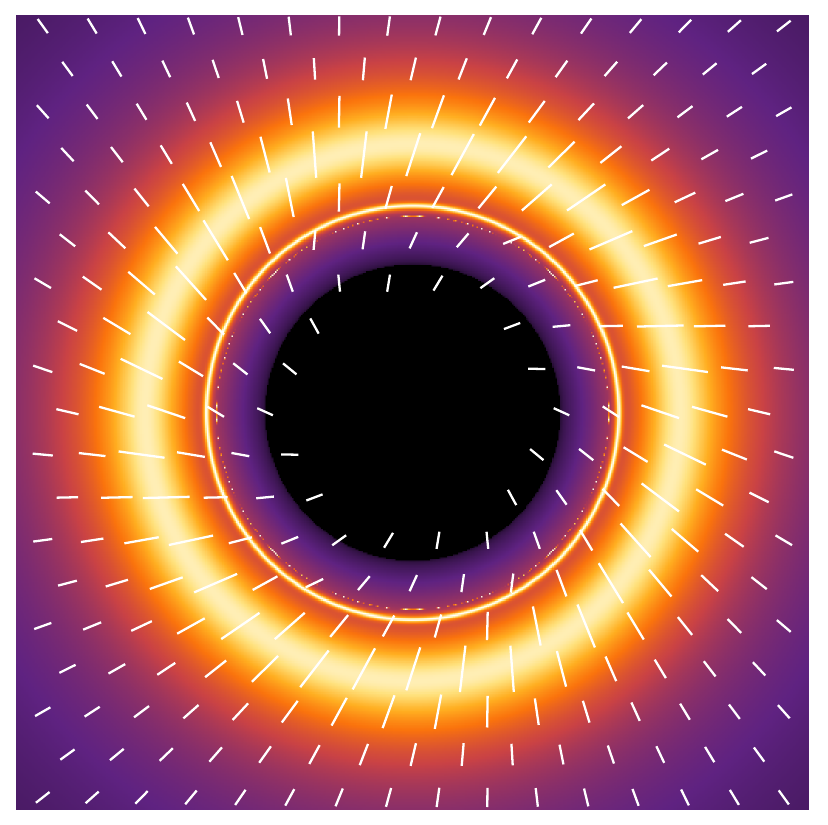}}
\subfigure[$(e)$ black hole with $\theta_{obs}=40^\circ$  ]{\includegraphics[width=.27\textwidth]{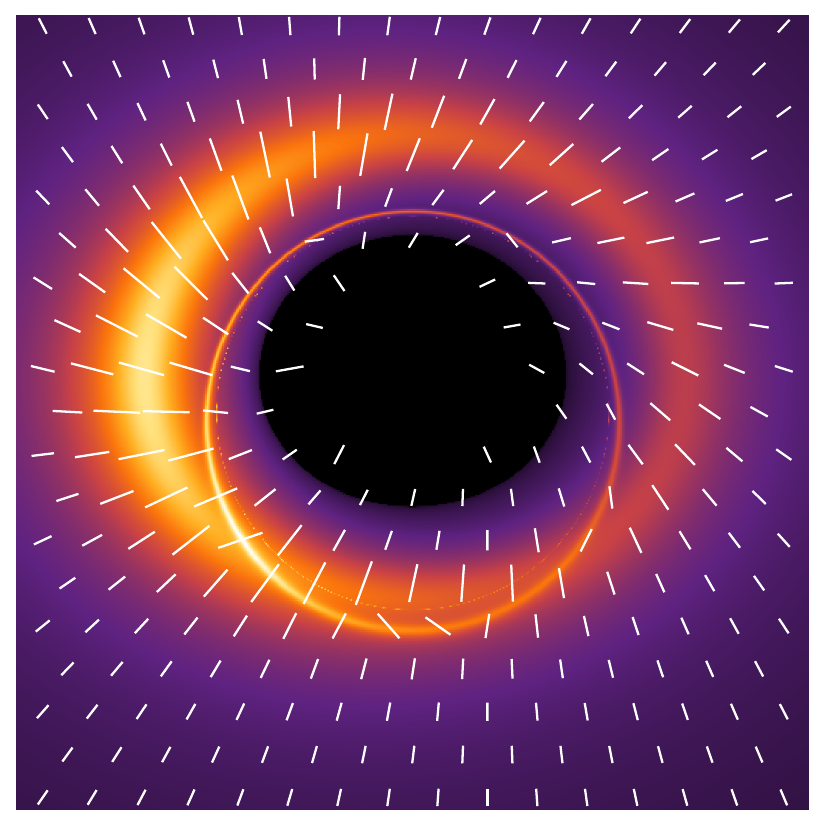}}
\subfigure[$(f)$ black hole with $\theta_{obs}=80^\circ$  ]{\includegraphics[width=.27\textwidth]{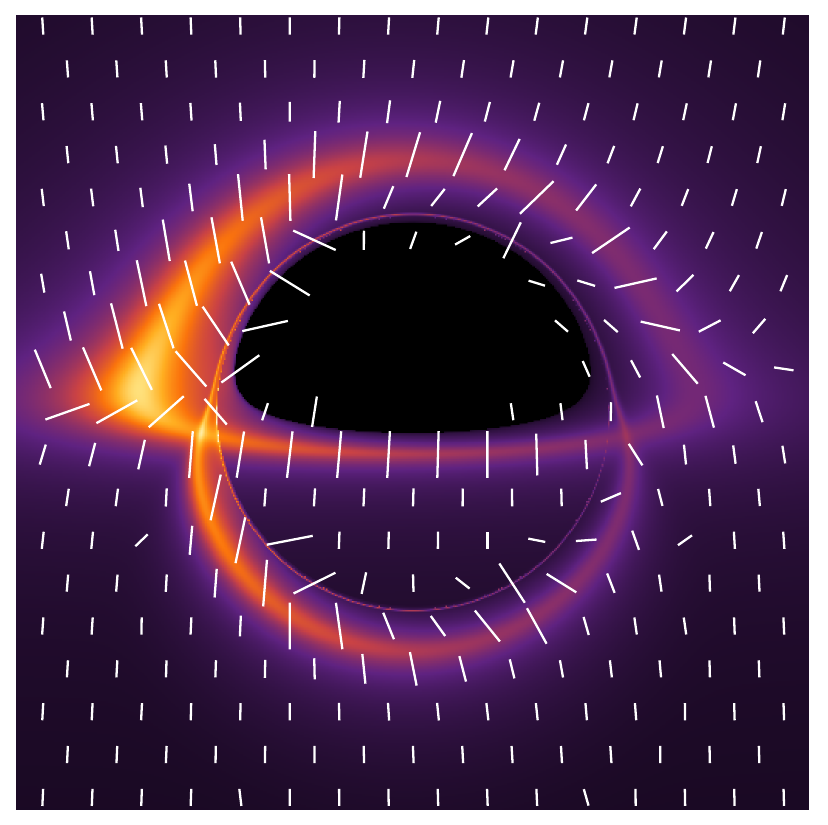}}
}
\vspace{-0.4cm}
\caption{\label{fig12} The images of boson star and Schwarzschild black hole.}
\end{figure}

In fact, we briefly analyzed the difference of the images between boson star and black hole in previous sections.
To establish a more intuitive comparison between the two scenarios, we further employ the { Model a} to numerically present the images of boson star and Schwarzschild black hole, as shown in Figure \ref{fig12}, where $\Lambda=100, \phi_0=0.15$ for boson star, and the observed positions are fixed as $r=200M, \theta_{obs}=0^\circ, 40^\circ, 80^\circ$, respectively.
\vspace{-0.1cm}

From the inside to outside of subfigure(a), the second light ring and its neighboring region represent the third-order image of the accretion disk, while the areas surrounding the first and third rings are lensed images. The brightest large ring corresponds to the direct image. As $\theta_{obs}=80^\circ$, the third-order image progressively forms into a ``horseshoe" shape.
In subfigure(d), the photon ring appears as the innermost first ring\footnote{An ultrathin, sharp light ring visible only upon image magnification.}. The region between the second luminous ring and the brightest large ring consists of lensed images, whereas the brightest large ring itself is the direct image. Meanwhile, the photon ring's shape and size remain invariant with the position of observer.
The direct image of the accretion disk in the spacetime of a boson star is consistent with that in the Schwarzschild black hole spacetime. However, more importantly, there are several distinct differences between the images of boson star and black hole:
(i) the contours of the lensed image of accretion disk differ significantly between the two, i.e., subfigures(c) and (f);
(ii) the higher-order images of the boson star exhibit a distinct ``crescent-shaped" light ring, which is absent in the Schwarzschild black hole;
(iii) black hole has a prominent inner shadow, whereas the boson star only exhibits a smaller, dimmer central region.
(iv) since both the lensed and higher-order images differ in size and shape, the polarization patterns between black holes and boson stars are very different. Notably, while the black hole's inner shadow region exhibits no polarization effect whatsoever, boson stars demonstrate significant polarization within the corresponding area.
In summary, the spacetime characteristics of boson star differ significantly from those of black hole, and these differences are precisely what lead to the variations in their observational images.
Based on the accretion disk images and polarized images,
it is evident that these significant observational features of boson stars that is absent in black holes may help us to effectively distinguish boson stars from black holes.

\begin{figure}[!h]
\makeatletter
\renewcommand{\@thesubfigure}{\hskip\subfiglabelskip}
\makeatother
\centering 
\vspace{-0.2cm}
\subfigure[$ $]{
\setcounter{subfigure}{0}
\subfigure[$(a)$ boson star with $\theta_{obs}= 0^\circ$]{\includegraphics[width=.25\textwidth]{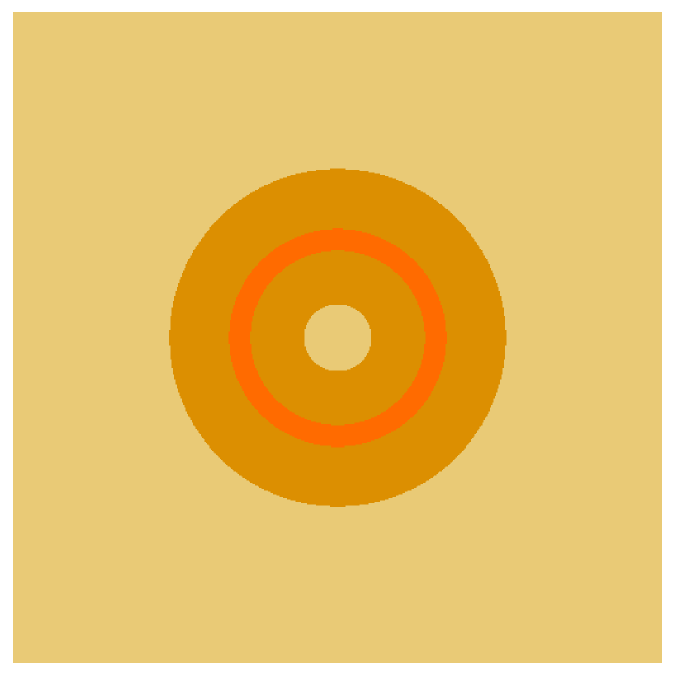}}
\subfigure[$(b)$ boson star with $\theta_{obs}= 80^\circ$]{\includegraphics[width=.25\textwidth]{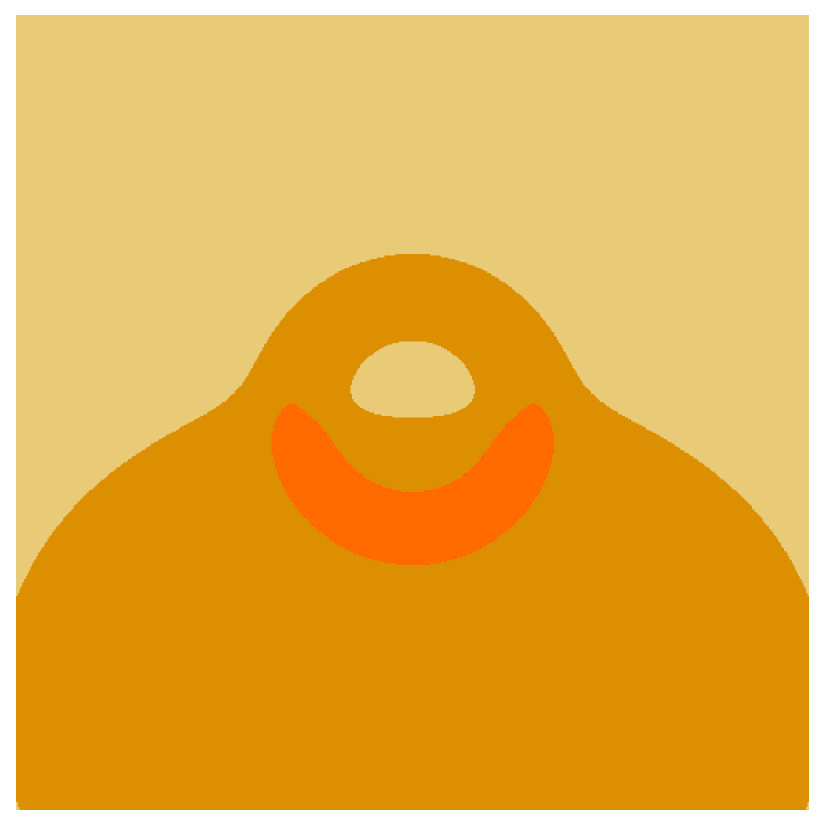}}
\subfigure[$(c)$ black hole with $\theta_{obs}=0^\circ$]{\includegraphics[width=.25\textwidth]{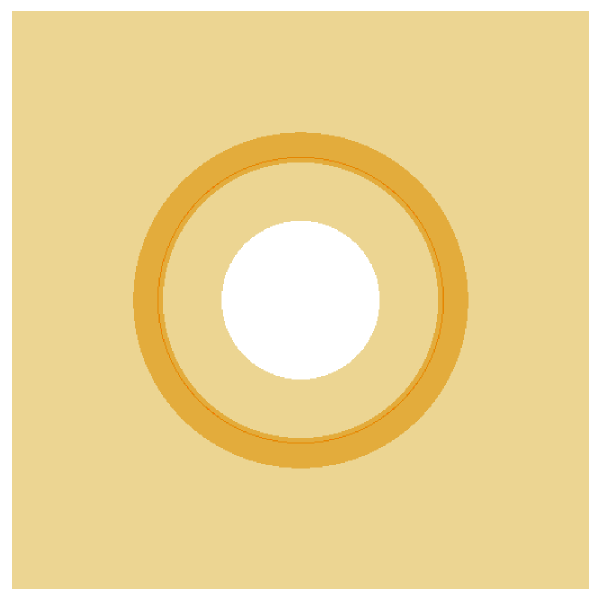}}
\subfigure[$(d)$ black hole with $ \theta_{obs}= 80^\circ$]{\includegraphics[width=.25\textwidth]{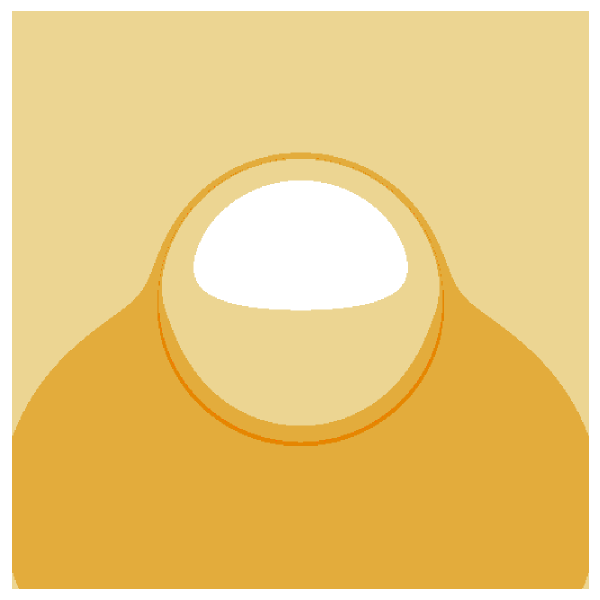}}
}
\vspace{-0.4cm}
\caption{\label{fig13} The number of intersections with the accretion disk.}
\end{figure}

To further investigate the space-time feature between boson stars and black holes, we present their observational images at $0^\circ$ and $80^\circ$, detailing the number of photon crossings through the thin disk, as shown in Figure \ref{fig13}.

\vspace{-0.1cm}

\begin{figure}[!h]
\makeatletter
\renewcommand{\@thesubfigure}{\hskip\subfiglabelskip}
\makeatother
\centering 
\vspace{-0.1cm}
\subfigure[$ $]{
\setcounter{subfigure}{0}
\subfigure[]{\includegraphics[width=.45\textwidth]{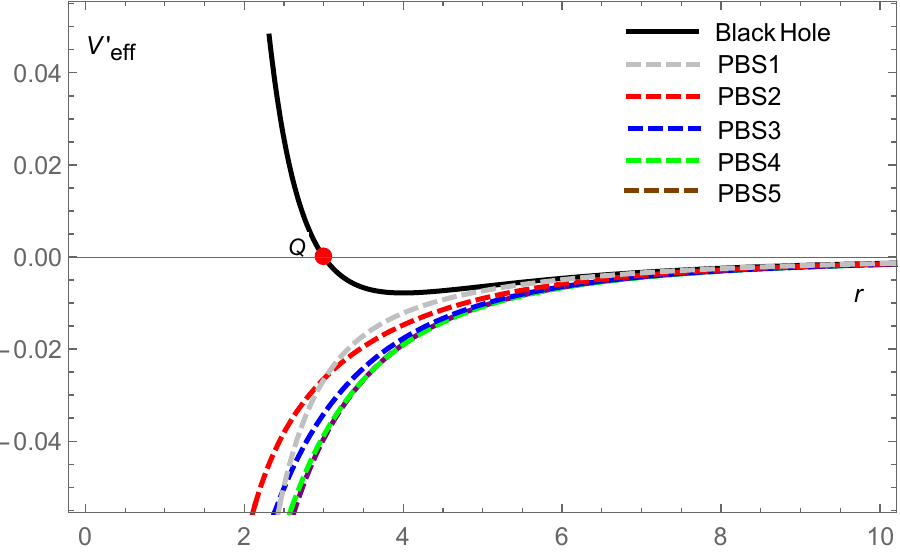}}}
\vspace{-0.5cm}
\caption{\label{fig14} Derivative of the effective potential.}
\end{figure}

For the subfigures (a) and (b) in Figure \ref{fig13}, the three colors (\raisebox{1\height}{\fcolorbox{black}{c1}{\rule{0.0cm}{0.0cm}}},
\raisebox{1\height}{\fcolorbox{black}{c2}{\rule{0.0cm}{0.0cm}}},
\raisebox{1\height}{\fcolorbox{black}{c3}{\rule{0.0cm}{0.0cm}}}) correspond to the cases where photons intersect with the accretion disk $n=1$, $n=2$, and $n=3$ times. For the subfigures (c) and (d), the four colors (\raisebox{1\height}{\fcolorbox{black}{c4}{\rule{0.0cm}{0.0cm}}},
\raisebox{1\height}{\fcolorbox{black}{c5}{\rule{0.0cm}{0.0cm}}},
\raisebox{1\height}{\fcolorbox{black}{c6}{\rule{0.0cm}{0.0cm}}},
\raisebox{1\height}{\fcolorbox{black}{c2}{\rule{0.0cm}{0.0cm}}}) correspond to the cases where photons intersect with the accretion disk $n=0$, $n=1$, $n=2$ and $n=3$ times.
It is evident that in the Schwarzschild black hole, the case of $n=0$ always exists, whereas it does not exist for boson stars. This indicates that, for a given field of view, when one traced light rays backward, all rays within the field of view of boson star will intersect the accretion disk. However, due to the existence of event horizon in Schwarzschild black hole, some light rays are directly absorbed by its horizon and do not intersect with the accretion disk.
More importantly, it can be seen from Figure \ref{fig13} that for boson star, the size and shape of the region where light rays intersect with the accretion disk three times constantly change with the observed angle($\theta_{obs}$). In contrast, the shape of this region remains almost unchanged for the black hole.
This means that the shape or position of higher-order images of boson stars(i.e., $n\geq3$) changes with $\theta_{obs}$, while this remains almost unchanged in the Schwarzschild black hole . For instance, when $n=3$, the image is always a circular shape and is consistently located at a fixed point $r_p$.
This seems to imply that while the photon ring exists in Schwarzschild black hole, no photon ring exists in boson star spacetime.
To further determine whether the photon ring exists in boson star spacetime, we presented the first derivative of potential function for both boson star and Schwarzschild black hole under the condition $V_{eff}=0$ in Figure \ref{fig14}, where $V_{eff}$ is the effective potential.

\vspace{-0.1cm}

From Figure \ref{fig14}, it can be observed that the first derivative of the potential function Schwarzschild black hole has a zero point $Q$ at $r=3M$ with $M=1$. This indicates that the potential function has an extremum, and the location of this point precisely corresponds to the position of photon ring of Schwarzschild black hole.
However, for boson stars, the first derivative of effective potential has no zero point. This implies that there are no photon rings in these boson star spacetimes.
As for the boson stars discussed in this paper, we can conclude that if the image of a compact object lacks a photon ring structure, then the compact object is definitely not a black hole, so it might possibly be a boson star.

\begin{figure}[!h]
\makeatletter
\renewcommand{\@thesubfigure}{\hskip\subfiglabelskip}
\makeatother
\centering 
\vspace{-0.2cm}
\subfigure[$ $]{
\setcounter{subfigure}{0}
\subfigure[$(a)$ Schwarzschild black hole]{\includegraphics[width=.25\textwidth]{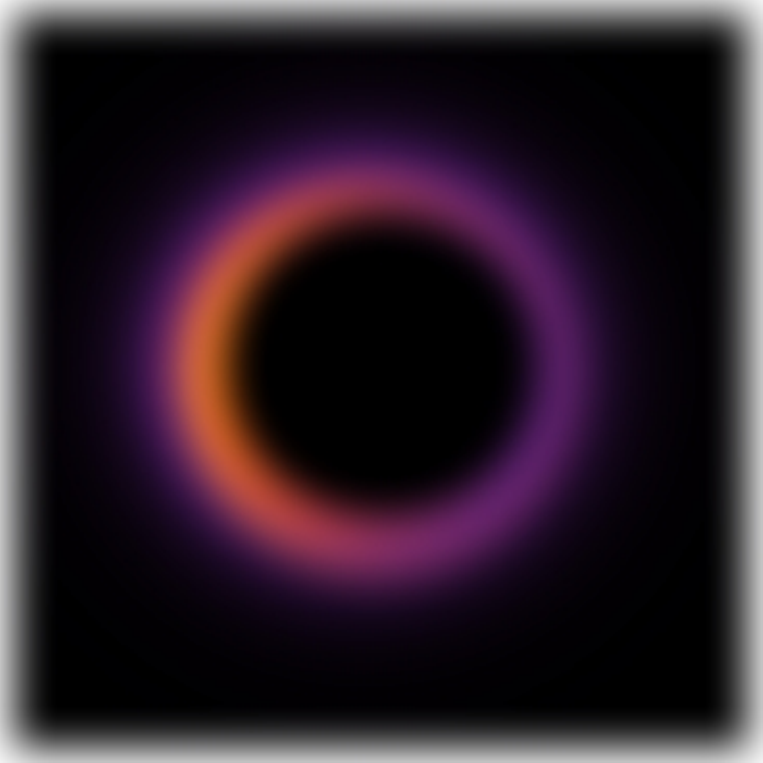}}
\subfigure[$(b)$ boson star for $\phi_0=0.15$]{\includegraphics[width=.25\textwidth]{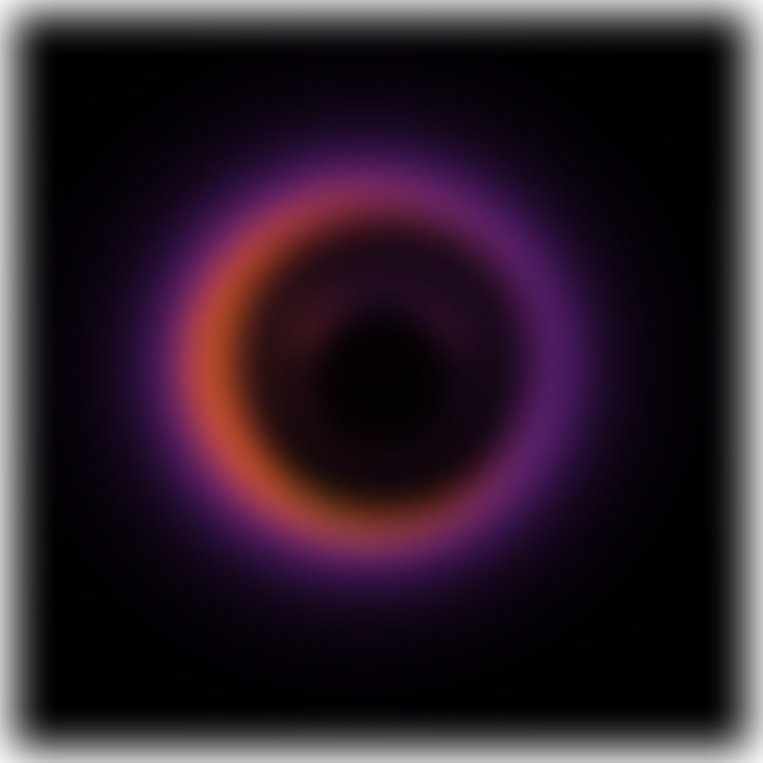}}
\subfigure[$(c)$ boson star for $\phi_0=0.06$]{\includegraphics[width=.25\textwidth]{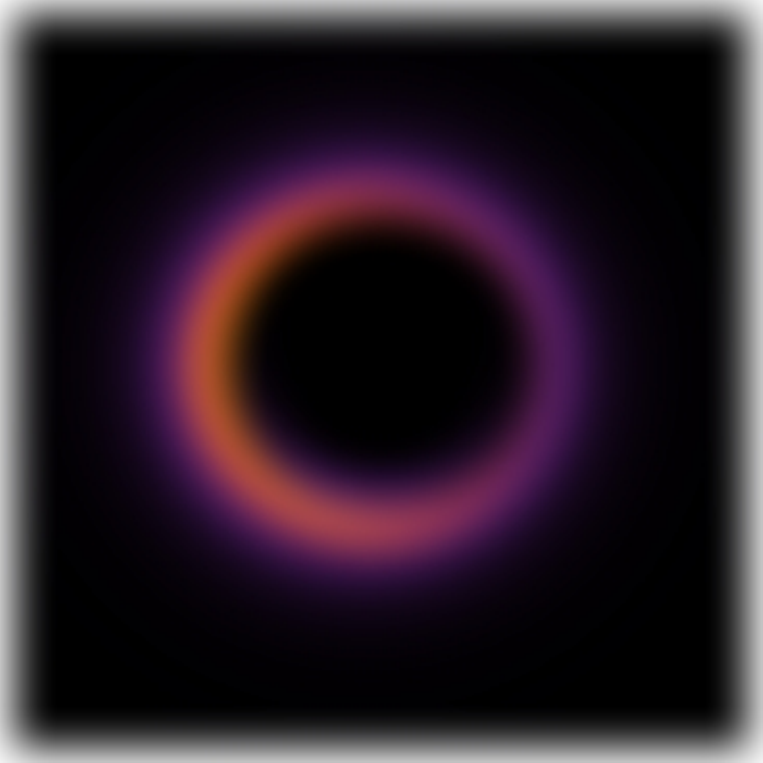}}
}
\subfigure[$ $]{
\setcounter{subfigure}{0}
\subfigure[$(d)$ Schwarzschild black hole]{\includegraphics[width=.25\textwidth]{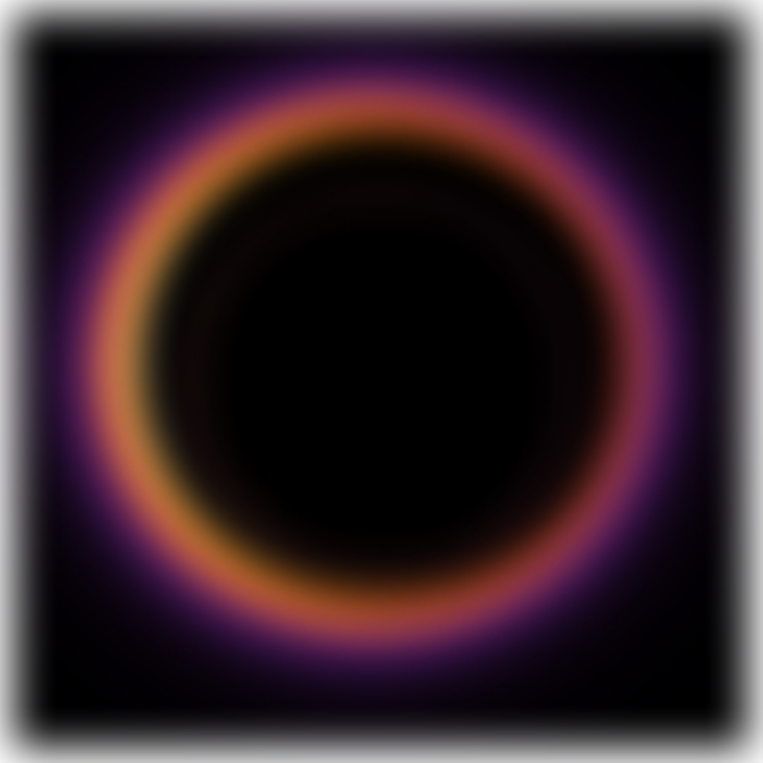}}
\subfigure[$(e)$ boson star for $\phi_0=0.15$]{\includegraphics[width=.25\textwidth]{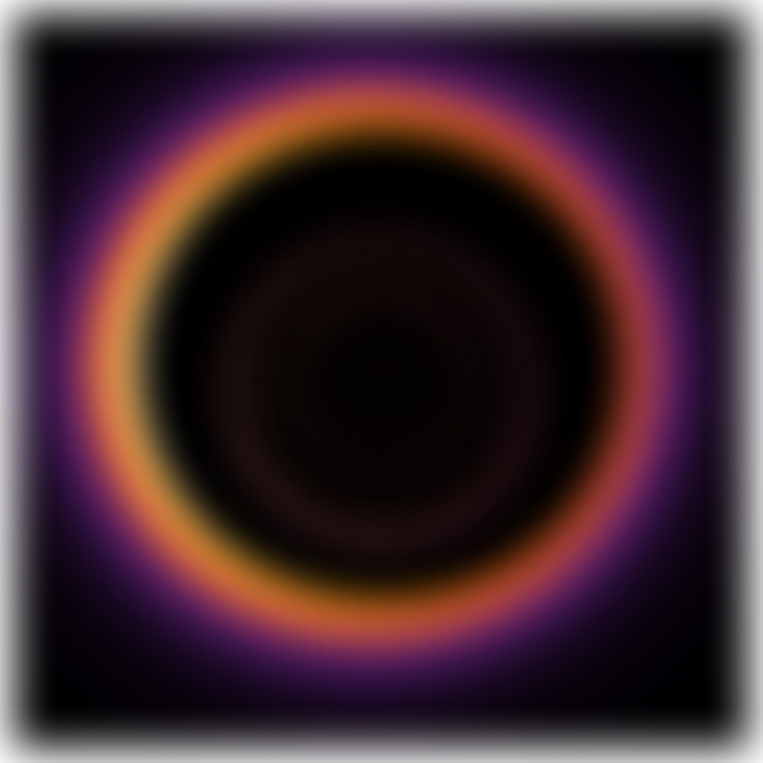}}
\subfigure[$(f)$ boson star for $\phi_0=0.06$]{\includegraphics[width=.25\textwidth]{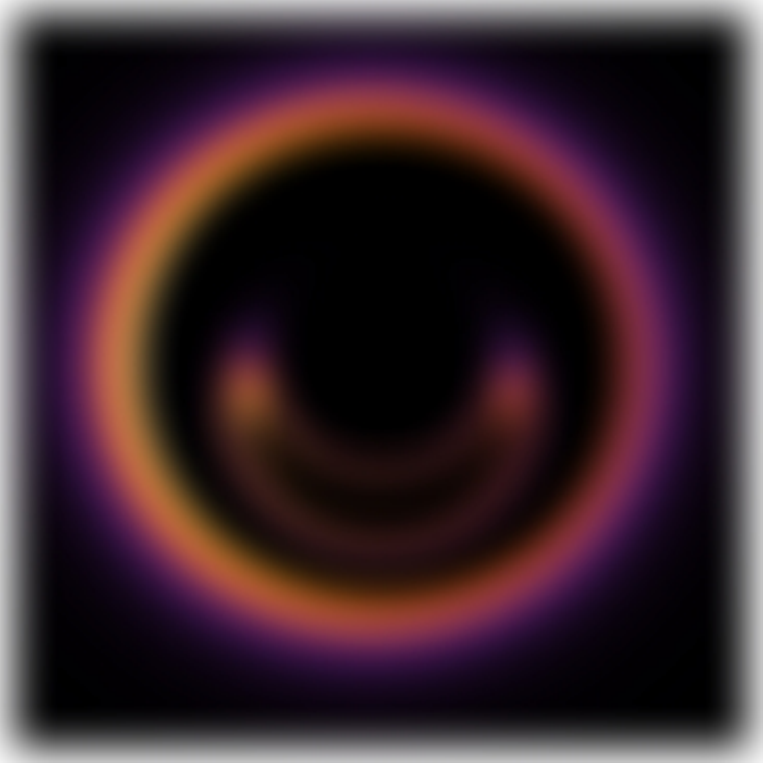}}
}
\vspace{-0.4cm}
\caption{\label{figblurry} The blurred images for black hole and boson stars with a Gaussian filter with a standard
deviation equal to $1/12$ of the field of view.}
\end{figure}

In summary, with a sufficient resolution, EHT could theoretically distinguish boson stars from black holes based on their thin-disk accretion images or polarized features. However, the current resolution of EHT remains inadequate for this purpose. We therefore analyze blurred images of both black hole and boson stars at present resolution.
In Figure\ref{figblurry}, we choose Schwarzschild black hole and two boson stars ($\phi_0=0.06,\phi_0=0.15$) as examples to show the blurred images at current resolution, where the accretion disk is fixed as Model b and $\Lambda = 200$.
\vspace{-0.1cm}

The first row of Figure \ref{figblurry} corresponds to the parameters $\beta=3.5M$, while the second row is $\beta=6M$, while the other parameters are $\sigma=1.25 M$, $\gamma=M$.
For subfigures (a)-(c), the inner boundaries of all accretion disks originate at $r=3.5M$. Although subfigure (b) exhibits some faint luminosity features within its ring structure, we still find it remarkably challenging to distinguish between the thin disk images of black hole and boson stars, which is particularly evident in figures (a) and (c).
When we position the inner edge of the thin disk at $6M$, the black hole remains nearly indistinguishable from the boson star($\phi_0=0.15$), while showing clearly distinguishable features when compared to the boson star($\phi_0=0.06$).
Therefore, we can conclude in this paper that, under the current EHT resolution, the accretion disk images of boson stars can indeed mimic or masquerade as black holes. Of course, under certain specific conditions, boson stars may still be observationally distinguished from black holes at current resolution.

\section{Conclusions and discussions}\label{sec6}

In this paper, based on the action for a complex scalar field minimally coupled to gravity, we numerically investigated the observable appearance of a class of boson stars under a thin accretion disk with the aid of the ray-tracing method.
In particular, employing a spherically symmetric spacetime background and a quartic-order self-interaction potential, we numerically obtained the solutions using the shooting method, where the condition that the asymptotic flatness at spatial infinity should be satisfied.
Further, we derive an analytical metric for these boson stars through a fitting approach. Finally, when a thin disk located at the equatorial plane, we have numerically obtained the thin disk images and polarized images of boson stars on the basis of the stereographic projection.

For different values of $\phi_0$ and $\Lambda$, we obtain ten specific boson stars.
As the coupling constant $\Lambda = 100$, the values of the central scalar field $\phi_0 = 0.06, 0.09, 0.12, 0.15, 0.18,$ correspond to the boson stars (PBS1, PBS2, PBS3, PBS4, PBS5) respectively, and for the case of $\phi_0 = 0.15$ with choice of $\Lambda =100, 200, 300, 400, 500,$ the related boson stars are labeled as ($\Lambda$BS6, $\Lambda$BS7, $\Lambda$BS8, $\Lambda$BS9, $\Lambda$BS10).
Through the metric fitting approach, one can see that the metric function $g_{tt}$ and $g_{rr}$ are always asymptotically flat and is well consistent with the Schwarzschild black hole for $r \to \infty$.
Then, by assuming the flows of thin disk moves along timelike circular orbits, the disk images of different boson stars are obtained in Figure \ref{figbp} for two Models a and b.
For the observer located at the position $\theta_{obs} = 17^{\circ}$, it shows for {Model a} that the number of light rings in those images increases as the parameter $\phi_0$ increases, that is, when $\phi_0 = 0.09$, the image only includes three light rings while the image shows five light rings for $\phi_0 = 0.15$ or $0.18$, with the positions and spacings of those rings being different.
In addition, due to the counterclockwise motion of the accretion flow, the intensity distribution of the image shows a distinct left-bright and right-dark pattern.
For Model b, when the accretion flows extends to the position $r=0$, the image exhibits a rather distinctive ``Central
Emission Region" pattern, i.e., the first two figures in third row of Figure \ref{figbp}. And, the black-hole-like shadow of boson stars no longer exist.
With the increase of parameters $\phi_0$ and $\Lambda$, the dark region in the boson star image becomes more pronounced, and the higher-order subrings in the center grow increasingly distinct.
Obviously, the size of the darker central region and lights rings also decreased with $\phi_0$ and $\Lambda$.
In section \ref{sec4}, for $\theta_{obs}=80^\circ$, one can not only easily distinguish the direct image, lensing image, and higher-order images of the boson star, but also observe the different features of the boson star¡¯s shape.
For example, by comparing with the case $\phi_0=0.06$, we can find that the image of case $\phi_0=0.15$ consists of four bright rings, each with distinct shapes and structural characteristics, which are Outer arc-shaped ring, Second outer symmetrical ring, Inner curved ring and Innermost small ring.

For the polarized features, the polarized intensity shows a positive correlation with the brightness of images of the thin disk, with significantly higher polarization in the high-brightness regions compared to the low-brightness regions. Therefore, since the intensity peak of the thin disk in Model b is located at the center of the boson star, while that of Model a is positioned away from the center, Model b exhibits a stronger polarized intensity than Model a at the center of the stars.
In addition, we find that the polarize intensity and direction show exhibit a very strong dependence on the coupling parameter $\Lambda$ and initial scalar field $\phi_0$, particularly in the central region of boson stars.
This implies that the polarized signatures of boson stars can effectively reveal the underlying spacetime geometry of it.
Also, when the position of the thin disk changes, both the intensity and direction polarized image in the central region of the boson star all exhibit some obvious variations, while its direction does not change with the position of the disk far away from the boson star.
So, we can see that the polarized image of boson stars is also closely related to the properties of thin disk and magnetic field.

On the other hand, compared with the black hole, it is obvious in boson stars that the contours of the lensed image of the thin disk differ significantly, and the distinct ``crescent-shaped" light ring is absent in a black hole.
And, within the event horizon of black holes, boson stars exhibit not only the image of the thin disk(i.e., ``Central Emission Region") but also the distinctive polarized effects, which are some remarkable phenomenon entirely absent in black hole systems.
The most important fact is that black hole has a photon ring, whereas the boson star obtained in this paper does not.
Combined with above facts, we can conclude that if the image of a compact object lacks a photon ring structure, then the compact object is definitely not a black hole, it might possibly be a boson star.
Therefore, if the future EHT observations detect the shadow and image of boson star, our research in this paper will effectively provide a possible basis for distinguishing the shadow and image of a black hole from that of a boson star.
And, compared to relying solely on accretion disk images, we emphasize that combining both the accretion disk images and polarized emission patterns of boson stars¡ªthen contrasting them with Schwarzschild black holes¡ªconstitutes a more robust and reliable approach to distinguishing between these two classes of compact objects.
In addition, under the current EHT resolution, we also find that the thin disk image of boson stars can indeed mimic or masquerade as black holes.
Obviously, this paper only focus on the thin accretion disk of massive boson stars. Further research on the thick disk of boson stars would also be a very interesting study, which may provide more reference information for distinguishing between boson stars and black holes.

\vspace{10pt}

\noindent {\bf Acknowledgments}
We would like to thank Minyong Guo and Jiewei Huang for their valuable discussions. This work is supported by the National Natural Science Foundation of China (Grant No.11903025), and by the Sichuan Science and Technology Program (2024NSFSC1999).

\end{document}